\documentclass[twocolumn,english,superscriptaddress,amsmath,floatfix,longbibliography,floats,prl]{revtex4-1}
\usepackage[colorlinks=true,urlcolor=blue,citecolor=blue,linkcolor=blue]{hyperref} 
\usepackage[T1]{fontenc}
\usepackage[utf8]{inputenc}
\usepackage{amssymb}
\usepackage{graphicx}
\usepackage{amsmath,color}
\usepackage{mathrsfs}
\usepackage{float}
\usepackage{indentfirst}
\usepackage{txfonts}
\usepackage[normalem]{ulem}
\usepackage[table]{xcolor}
\usepackage{array,ragged2e}
\usepackage{multirow}
\usepackage{grffile}
\makeatletter

\newcommand{\ave}[1]{\langle #1 \rangle}

\makeatother
\usepackage{babel}
\begin{document}
\hyphenpenalty=5000
\tolerance=1000
\title{Emergent Symmetry in Quantum Phase Transition: From Deconfined Quantum Critical Point to Gapless Quantum Spin Liquid}
\author{Wen-Yuan Liu}
 \affiliation{Department of Physics, The Chinese University of Hong Kong, Shatin, New Territories, Hong Kong, China}
 \affiliation{Division of Chemistry and Chemical Engineering, California Institute of Technology, Pasadena, California 91125, USA}
 \author{Shou-Shu Gong}
 \affiliation{Department of Physics, Beihang University, Beijing 100191, China}
 \author{Wei-Qiang Chen}
 \affiliation{Shenzhen Institute for Quantum Science and Engineering and Department of Physics, Southern University of Science and Technology, Shenzhen 518055, China}
 \affiliation{Shenzhen Key Laboratory of Advanced Quantum Functional Materials and Devices, Southern University of Science and Technology, Shenzhen 518055, China}
 \author{Zheng-Cheng Gu}
 \affiliation{Department of Physics, The Chinese University of Hong Kong, Shatin, New Territories, Hong Kong, China}
\date{\today }
\begin{abstract}

\noindent\textbf{Abstract}\\
The emergence of exotic quantum phenomena in frustrated magnets is rapidly driving the development of quantum many-body physics, raising fundamental questions on the nature of quantum phase transitions. 
Here we unveil the behaviour of emergent symmetry involving two extraordinarily representative phenomena, i.e., the deconfined quantum critical point (DQCP) and the quantum spin liquid (QSL) state. Via large-scale tensor network simulations, we study a spatially anisotropic spin-1/2 square-lattice frustrated antiferromagnetic (AFM) model, namely the $J_{1x}$-$J_{1y}$-$J_2$ model, which contains anisotropic nearest-neighbor couplings $J_{1x}$, $J_{1y}$ and the next nearest neighbor coupling $J_2$. For small $J_{1y}/J_{1x}$, by tuning $J_2$, a direct continuous transition between the AFM and valence bond solid  phase is observed.(Of course, the possibility of weakly first order transition can not be fully excluded.) With growing $J_{1y}/J_{1x}$, a gapless QSL phase gradually emerges between the AFM and VBS phases. 
We observe an emergent O(4) symmetry along the AFM--VBS transition line, which is consistent with the prediction of DQCP theory. 
Most surprisingly, we find that such an emergent O(4) symmetry holds for the whole QSL--VBS transition line as well.  These findings reveal the intrinsic relationship between the QSL and DQCP from categorical symmetry point of view, and strongly constrain the quantum field theory description of the QSL phase. The phase diagram and critical exponents presented in this paper are of direct relevance to future experiments on frustrated magnets and cold atom systems.

\end{abstract}
\maketitle
\noindent\textbf{Introduction}\\
The concept of deconfined quantum critical point (DQCP) was proposed two decades ago to describe Landau forbidden continuous phase transitions between two ordered phases, such as the transition between the antiferromagnetic (AFM) and valence bond solid (VBS) phases~\cite{DQCP1,DQCP2}. Since then, the DQCP has been investigated in a number of numerical studies on various spin, fermion, and classical loop models~\cite{JQ2007,JQ2008,JQ2008_2,JQ2009,loopmodel2,charrier2010,JQ2010,JQ2011,JQ2013,JQ2013_2,JQ2013_3,JQ2015,loopmodel1,JQ2016,sreejith2019,fermionDQCP2016,fermionDQCP2016,fermionDQCP2017,fermionDQCP2018,fermionDQCP2018_2,fermionDQCP2019,liuQSL,liuj1j2j3}. One of the most remarkable discoveries is the appearance of enhanced symmetry~\cite{loopmodel2,sreejith2019}, which  is essential for understanding the underlying physics. However, in various DQCP-related studies, unusual scaling violation has been observed and the expected continuous nature of the transition has been challenged by the possibility of weakly first-order transition. Such perplexing phenomena raise a puzzle regarding the nature of the DQCP.  

In a recent breakthrough, the intrinsic relationship between the DQCP and gapless quantum spin liquid (QSL) was revealed that a gapless QSL phase can develop from a DQCP~\cite{liuj1j2j3}. This demonstrates a new perspective to understand both DQCP and QSL, implying that they could be described by a unified quantum field theory. However, as the DQCP physics highly depends on  microscopic symmetry such as spin and lattice symmetry~\cite{senthil2006,loopmodel2, block2013,assaad2017,qin2017,wangchong2017,zhang2018,nahum2019,sreejith2019,shyta2022}, it is an open question whether this is a generic relation for systems with different symmetries. 
In particular, the behavior of emergent symmetry is a critical concern and a fundamental aspect in developing a quantum field theory description. 

On the other hand, the categorical symmetry framework and holographic principle suggest that emergent symmetry may exist for generic quantum phase transitions beyond the Landau paradigm~\cite{catsymmetry1}. While some examples have been explored for one-dimensional systems~\cite{catsymmetry2}, it is still uncertain which types of quantum phase transitions in higher dimensions support emergent symmetry. 
Specifically, it is unknown whether the quantum phase transition into a gapless QSL also exhibits emergent symmetry or not.


Here we present an invaluable scenario that significantly enhances our understanding of both DQCP and QSL with $C_2$ lattice systems. Starting with the DQCP-type AFM-VBS transition, by tuning the coupling constants, a gapless QSL phase emerges in between the AFM and VBS phases. Most  surprisingly, we observe that the emergent symmetry arises not only at the DQCP but also persists at the phase boundary of the QSL-VBS transition. 
Since the quantum phase transition at the phase boundary of a gapless QSL is unlikely to be first order, we believe that the corresponding emergent O(4) symmetry should survive even in the thermodynamic limit. These findings shed new light on the intrinsic relation between DQCP and gapless QSL from the categorical symmetry perspective.

\bigskip
\noindent\textbf{Results}\\ 
\noindent\textbf{Model.}\\
We focus on the rectangular spin-1/2 model, the frustrated $J_{1x}$-$J_{1y}$-$J_2$ model~\cite{Tsvelik2003,balents2004}, which contains anisotropic nearest-neighbor AFM Heisenberg couplings $J_{1x}>0,J_{1y}>0$ and the next nearest neighbor AFM Heisenberg coupling $J_{2}>0$, with the Hamiltonian:
\begin{equation}
H=J_{1x}\sum_{\langle i,j \rangle_x}\mathbf{S_i}\cdot\mathbf{S_j}+J_{1y}\sum_{\langle i,j \rangle_y}\mathbf{S_i}\cdot\mathbf{S_j} +J_2\sum_{\langle\langle
i,j\rangle\rangle}\mathbf{S_i}\cdot\mathbf{S_j}.
\label{model}
\end{equation}
This model was introduced to study the interplay between quantum frustration and spinon excitations~\cite{Tsvelik2003}. The strong frustration present in this model makes it challenging to simulate accurately, and thus its global phase diagram remains elusive, despite previous studies~\cite{balents2004,Sindzingre2004,Bishop2008}. 

\begin{figure}[tbp]
 \centering
 \includegraphics[width=3.3in]{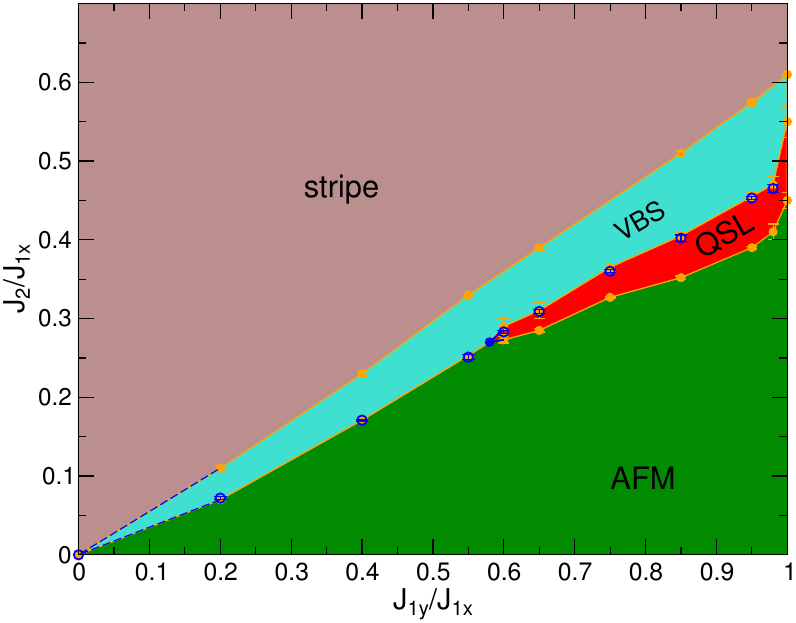}
 \caption{Ground-state phase diagram of the $J_{1x}$-$J_{1y}$-$J_2$ model, including four phases: the AFM, VBS, and gapless QSL (red region) phases, and a stripe phase. The dashed blue lines denote the hypothetical shape of the VBS phase close to the origin. Solid blue lines in the middle region denote the unknown QSL shape close to the tricritical point (filled blue circle). Open blue circles have emergent O(4) symmetry.}
 \label{fig:J1xJ1yJ2phaseDiagram}
 \end{figure}

Recently, the advancement in tensor network methods, specifically the finite projected entangled pair state algorithm ~\cite{liu2017,liufinitePEPS}, has provided a powerful tool for investigating frustrated models with high accuracy~\cite{liuQSL,liuj1j2j3}
 By applying such a state-of-the-art method, we elaborately investigated this model through performing large-scale computations. The global phase diagram is shown in Fig.~\ref{fig:J1xJ1yJ2phaseDiagram}. In the small $J_{1y}$ region, we observe a direct AFM--VBS transition with an emergent O(4) symmetry, formed by three-component AFM order parameters and the one-component VBS order parameter. In the larger $J_{1y}$ region, we observe a gapless quantum spin liquid (QSL) phase between the AFM and VBS phases. Surprisingly, the emergent O(4) symmetry persistently exists on the QSL--VBS transition line.

\begin{figure}[htbp]
 \centering
 \includegraphics[width=3.4in]{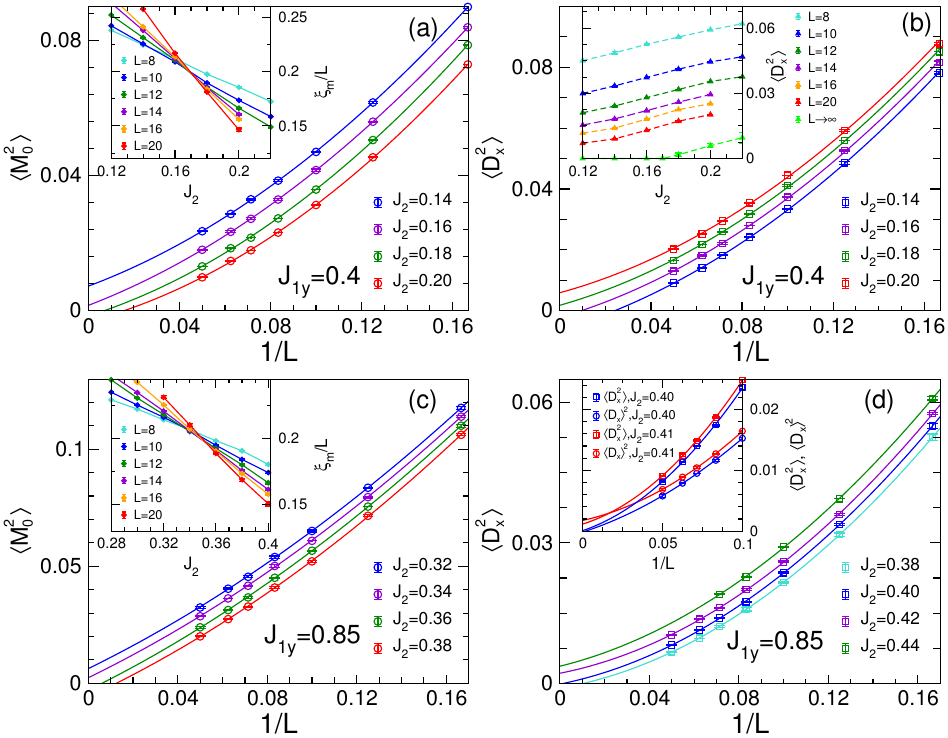}
 \caption{(a) Finite size scaling of the AFM order parameter (main panel)  and crossing of $\xi_m/L$ (inset) at $J_{1y}=0.4$. (b) Finite size scaling of the VBS order parameter (main panel) and  $J_2$-dependence of VBS order parameters at $J_{1y}=0.4$. (c) Finite size scaling of the AFM order parameter (main panel)  and crossing of $\xi_m/L$ (inset) at $J_{1y}=0.85$. (d) Finite size scaling of VBS order parameters including  $\langle D^2_x \rangle$ and boundary-induced dimerization $\langle D_x \rangle^2$ at $J_{1y}=0.85$. Second-order polynomial fits are used for all extrapolations.} 
 \label{fig:OrderParameterJ1y_04_085}
 \end{figure}

  \bigskip
 \noindent\textbf{Continuous AFM-to-VBS transition.} We set $J_{1x}=1$ throughout the paper and sweep $J_2$ with fixed $J_{1y}$ to obtain the phase diagram.  We first consider the 
large anisotropy region, where we find that a direct AFM--VBS transition can occur up to $J_{1y}=0.55$ but probably vanishes at $J_{1y} \simeq 0.6$. The AFM order parameter $\ave{M^2_0}$ is defined as the spin order parameter $m^2({\bf k})=\frac{1}{L^4}\sum_{\bf{ij}}\langle{\bf S}_{{\bf i}}\cdot {\bf S}_{{\bf j}}\rangle {e}^{i {\bf k}\cdot({\bf i}-{\bf j})}$ at ${\bf k}=(\pi,\pi)$, where ${\bf i}=(i_x,i_y)$ is the site position.  Taking $J_{1y}=0.4$ as an example, we show the AFM order parameter on different $L\times L$ systems up to $20\times 20$ in Fig.~\ref{fig:OrderParameterJ1y_04_085}(a). The finite size scaling of the system suggests that the AFM order vanishes at  $J_{c1}=J_2\simeq 0.17$ in the two-dimensional (2D) limit. We also use the crossing of the dimensionless quantity $\xi_m/L$ to determine the transition point, where $\xi_m$ is the spin correlation length defined as $\xi_m=\frac{L}{2\pi}\sqrt{\frac{m^2(\pi,\pi)}{m^2 (\pi+2\pi/L,\pi)}-1}$~\cite{liuQSL}. This gives rise to a consistent $J_{c1}$, as shown in the inset of Fig.~\ref{fig:OrderParameterJ1y_04_085}(a).

The dimer order parameter $D_{\alpha}=\frac{1}{N_b}\sum_{{\bf i}}(-1)^{i_{\alpha}} B^{\alpha}_{\bf i}$ is used to detect possible VBS patterns,  where $B^{\alpha}_{\bf i}={\bf S}_{{\bf i}} \cdot {\bf S}_{{\bf i}+{\rm e_\alpha}}$ is the bond operator between nearest sites  ${\bf i}$ and ${\bf i}+{\rm {\bf e}_\alpha}$  with $\alpha=x$ or $y$, and $N_b=L(L-1)$ is the total number of counted bonds along the $\alpha$ direction for open-boundary systems.  Fig.~\ref{fig:OrderParameterJ1y_04_085}(b) presents the horizontal VBS order parameter  $\langle D^2_x \rangle$  with the largest system size up to a $20\times 20$ matrix at fixed $J_{1y}=0.4$. It is seen that the extrapolated value of $\langle D^2_x \rangle$  for the 2D limit is zero at $J_2=0.16$ but nonzero at $J_2=0.18$. Note that the $y$-direction VBS order parameter $\langle D^2_y \rangle$ is very small for finite sizes and clearly extrapolates to zero in the 2D limit. The results indicate that the VBS order sets in at $J_{c2}=J_2\simeq 0.17$, and there is thus a direct AFM--VBS transition at $J_c=J_{c1}=J_{c2}$. We later confirm such an AFM--VBS transition through other means.
The order parameters for each system size have a smooth change with $J_2$, as presented in the inset of Fig.~\ref{fig:OrderParameterJ1y_04_085}(b), and the AFM--VBS transition is thus likely to be continuous, although the possibility of a weakly first-order transition cannot be fully excluded.
 
\bigskip
\noindent\textbf{Emergence of the QSL phase.} For $J_{1y}\geq 0.6$, we do not find a direct transition between the AFM and VBS phases, and instead, a QSL phase develops in between. Taking $J_{1y}=0.85$ as an example, we present the AFM order parameter in Fig.~\ref{fig:OrderParameterJ1y_04_085}(c). The finite-size scaling of the AFM order parameter at different $J_2$ suggests that the AFM order begins to vanish at $J_{c1}=J_2\simeq 0.35$ in the 2D limit, which is further supported by the crossing of $\xi_m/L$. The horizontal dimer order parameter $\langle D^2_x \rangle$ in the 2D limit develops above $J_{2}\simeq 0.4$, as seen in Fig.~\ref{fig:OrderParameterJ1y_04_085}(d). We also examine the dimerization $\ave{D_x}^2$ induced by open boundaries as a further check. As shown in the inset of Fig.~\ref{fig:OrderParameterJ1y_04_085}(d), the extrapolated values at $J_{2}=0.4$ are zero whereas those at $J_{2}=0.41$ are 0.0012(7) for $\langle D^2_x \rangle$ and 0.0018(3) for $\langle D_x \rangle^2$. The results consistently suggest the onset of the VBS order at $J_{c2} \simeq 0.405(5)$ and indicate a QSL phase for $0.35\lesssim J_2 \lesssim 0.4$ by excluding spin and dimer orders. The calculations for other $J_{1y}$ up to $J_{1y}=0.98$ are shown in the Supplemental Information. The global phase diagram is presented as Fig.~\ref{fig:J1xJ1yJ2phaseDiagram} and shows that a (gapless) QSL phase can develop from a DQCP.
\begin{figure}[htbp]
 \centering
 \includegraphics[width=3.4in]{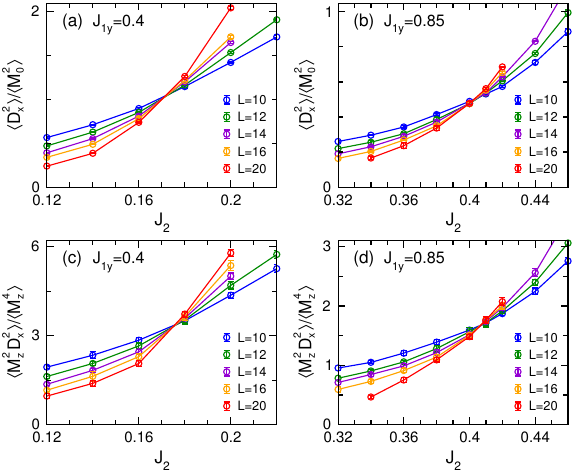}
 \caption{Ratios of $\langle  D^2_x \rangle / \langle M^2_0\rangle$ (a,b) and $\langle M^2_z D^2_x \rangle / \langle M^4_z\rangle$ (c,d) for $J_{1y}=0.4$ and $J_{1y}=0.85$ for the verification of emergent symmetry. The crossing $J_2$ values of $\langle  D^2_x \rangle / \langle M^2_0\rangle$ and $\langle M^2_z D^2_x \rangle / \langle M^4_z\rangle$ are approximately 0.171 and 0.176  for $J_{1y}=0.4$ and  $0.402$ and $0.407$ for $J_{1y}=0.85$, respectively.}
 \label{fig:O4_higherMoment}
 \end{figure}

\bigskip
 \noindent\textbf{Emergent O(4) symmetry.} For the rectangular anisotropy case that we consider here, it has been argued that O(4) symmetry emerges at the AFM--VBS transition point~\cite{metlitski2018} through the rotation of the three-component AFM vector ${\bf M}=(M_x, M_y, M_z)$ and one-component VBS order parameter $D_x$ into each other to form a superspin: ${\bf n}=({ M_x, M_y, M_z, D_x } )$. According to Refs.~\cite{loopmodel2,sreejith2019,nahum2019}, if O(4) symmetry emerges, the moments of the order parameter should satisfy certain relations. Once the SO(3) symmetry acting on the AFM vector ${\bf M}=(M_x, M_y, M_z)$ is satisfied, it is sufficient to demonstrate the fully emergent O(4) symmetry by verifying an additional emergent symmetry rotating $M_z$ into $D_x$. In our calculation, we confirm the good SO(3) symmetry of the ground state around the transition point and inside the nonmagnetic phases, where each spin component is given as  $\ave{M^2_{\alpha}}=\frac{1}{3}\ave{M^2_0}$ with $\ave{M^2_0}$ being the AFM order parameter. 
 
Once SO(3) symmetry is satisfied, we only need to check the additional symmetry formed by $M_z$ and $D_x$.
A simple but nontrivial quantity of the emergent O(4) symmetry is that at the transition point $J_c$, the ratio between the order parameters $\langle D^2_x \rangle / \langle M^2_0\rangle$ should be independent of the system size ~\cite{loopmodel2,sreejith2019,nahum2019}. 
In Fig.~\ref{fig:O4_higherMoment}(a) for $J_{1y}=0.4$, we present the order parameter ratio for different system sizes at different $J_2$, which give rise to almost the same crossing $J_2$, with the phase transition point determined from the crossing of $\xi_m/L$ and finite size scaling of order parameters. We also consider the higher-order moments of order parameters~\cite{loopmodel2,sreejith2019,nahum2019}, which are challenging to compute. Within our capability, we compute $\langle M^4_z \rangle$ and  $\langle M^2_z D^2_x \rangle$ to verify whether the crossing of the ratio $\langle M^2_z D^2_x \rangle/\langle M^4_z \rangle$ is located at the transition point $J_c$. As shown in Fig.~\ref{fig:O4_higherMoment}(c) for $J_{1y}=0.4$, the crossing $J_2$ value is indeed in good agreement with $J_c$ determined in other ways. These results strongly support the emergence of O(4) symmetry at the AFM--VBS transition point. Results supporting the emergent O(4) symmetry at the AFM--VBS transition points for fixed $J_{1y}=0.2$ and 0.55 can be found in the Supplemental Information.

We now move to the weak-anisotropy region where QSL appears. In this situation, in contrast with the AFM--VBS transition, we find that the crossing $J_2$ values of $\langle D^2_x \rangle / \langle M^2_0\rangle$ in the 2D limit are different from those of $\xi_m /L$. Within our resolution, for each $J_{1y}$, we find that the crossing $J_2$ values of $\langle D^2_x \rangle / \langle M^2_0\rangle$ are almost the same as the QSL--VBS transition points obtained by the finite size scaling of VBS order parameters [Fig.~\ref{fig:O4_higherMoment}(b)]. These values are listed in Table ~\ref{tab:criticalpoint} for convenient comparison. Usually, the crossings of $\langle D^2_x \rangle / \langle M^2_0\rangle$ have much smaller finite effects, as has been observed in other DQCP studies~\cite{loopmodel2,sreejith2019}. The crossings of $\xi_m /L$ are somewhat shifted for small systems but seem to almost converge at large system sizes up to $20\times 20$, and we adopt collective fitting to collapse the data in accounting for the finite-size effects (see more results in the Supplemental Information).
   \begin{table}[htbp]
   \centering
 \caption {Phase transition points for different quantities and a fixed $J_{1y}$. The second column gives the transition points $J_{c1}$ for the AFM--VBS or AFM--QSL transition obtained by the collective fitting of $\xi_m /L$. The third column gives the estimated crossing $J_2$ values of $\langle D^2_x \rangle / \langle M^2_0\rangle$ for $L\rightarrow \infty$. The last column gives the estimated AFM--VBS or QSL--VBS transition point from the finite size scaling (FSS) of VBS order parameters.}
	\begin{tabular*}{\hsize}{@{}@{\extracolsep{\fill}}cccl@{}}
		\hline\hline
	      $J_{1y}$ &  $\xi_m/L$ & $\langle D^2_x\rangle /\langle M^2_0\rangle$ & FSS (VBS)    \\ \hline
 		0.20   & 0.070(2)   & 0.071(2) & 0.07(1) \\
 		0.40   & 0.171(3)   & 0.171(2)& 0.17(1)  \\
 		0.55  & 0.253(2)   & 0.255(2) & 0.25(1)   \\
 		0.60   & 0.273(4) & 0.283(1) & 0.29(1)   \\
 		0.65  & 0.285(2) & 0.309(1) & 0.31(1)   \\
 		0.75  & 0.327(2) & 0.360(1) & 0.365(5)   \\
 		0.85  & 0.352(5) & 0.402(3) & 0.405(5)  \\
 	    0.95  & 0.390(2) & 0.453(2) &0.455(5)   \\
 	    0.98  & 0.410(5) & 0.465(3) & 0.47(1)  \\
 		\hline\hline
	\end{tabular*}
\label{tab:criticalpoint}	
\end{table}
The crossing $J_2$ values of $\xi_m/L$ and $\langle D^2_x \rangle / \langle M^2_0\rangle$ coincide well in the strong-anisotropy region (for example, $J_{1y}=0.2,0.4$, and $0.55$) but disagree in the weak-anisotropy region ($J_{1y}=0.6,0.65,0.75,0.85,0.95$, and $0.98$), which is strong evidence that in between the AFM and VBS phases for $J_{1y}\gtrsim 0.6$ there exists an intermediate phase, namely the QSL. More interestingly, the coincidence of crossings from $\langle D^2_x \rangle / \langle M^2_0\rangle$ and QSL--VBS transition points indicates the emergence of O(4) symmetry on the QSL--VBS phase boundary. We compute the four order moments of the order parameter at $J_{1y}=0.85$ [Fig.~\ref{fig:O4_higherMoment}(d)], which gives rise to almost the same crossing as that of $\langle D^2_x \rangle / \langle M^2_0\rangle$ and further supports the emergent O(4) symmetry.

Note that in the QSL phase,  $\langle M^2_0\rangle$ is zero in the thermodynamic limit, but for a finite size, the ratio $\langle D^2_x \rangle / \langle M^2_0\rangle$  is still meaningful. The QSL is gapless with a power-law decay for both dimer and spin correlation functions, and the ratio $\langle D^2_x \rangle / \langle M^2_0\rangle$ for finite system sizes thus reflects the relative decay rate $L^{-(\alpha_d-\alpha_s)}$ assuming $\langle D^2_x \rangle \propto L^{-\alpha_d}$ and $\langle M^2_0 \rangle \propto L^{-\alpha_s}$. At the QSL--VBS transition point, the size-independent $\langle D^2_x \rangle / \langle M^2_0\rangle$ indicates that spin and dimer correlations both decay algebraically with the same exponents, which will be verified in the next section. We note that the dominant spin correlation in the QSL phase is still AFM; i.e., the peak of the spin structure factor is at ${\bf k_0}=(\pi,\pi)$. This explains why $\langle M^2_0\rangle$ should be used instead of other ${\bf k}-$value magnetic moments. 
\begin{figure*}[htbp]
 \centering
 \includegraphics[width=6.8in]{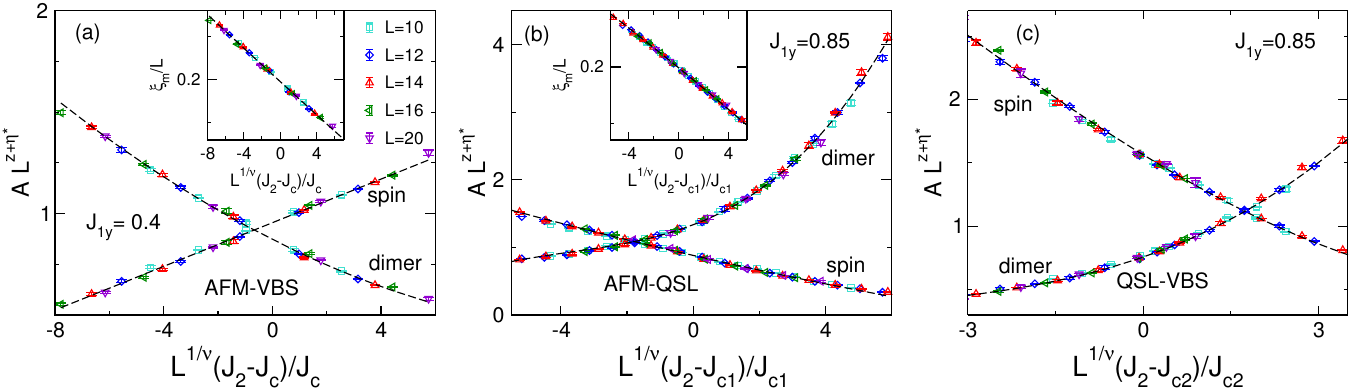}
 \caption{Scaling analysis of physical quantities for AFM and VBS order parameters (main panels), and the spin correlation length (insets): (a) at the AFM--VBS transition point with fixed $J_{1y}=0.4$, using $\nu=0.85$, $J_c=0.171$, and $z+\eta_s^*=z+\eta_d^*=1.37$; (b) at the AFM--QSL transition point with fixed $J_{1y}=0.85$, using $\nu=1.0$, $J_{c1}=0.352$, $z+\eta_s^*=1.18$, and $z+\eta_d^*=1.85$; and (c) at the QSL--VBS transition point with fixed $J_{1y}=0.85$, using $\nu=1.0$, $J_c=0.402$, and $z+\eta_s^*=z+\eta_d^*=1.51$.  Subleading corrections are not used in these cases. Black dashed lines are quadratic curves drawn using corresponding critical exponents. }
 \label{fig:DataCollapse}
 \end{figure*}
 
\bigskip
 \noindent\textbf{Critical exponents.} We extract the critical exponents of the AFM--VBS, AFM--QSL and QSL--VBS transitions to further analyze these unconventional phase transitions (see the Supplemental Information for more details). Figure~\ref{fig:DataCollapse} shows the data collapse of physical quantities including $\ave{M^2_0}$, $\ave{D^2_x}$, and $\xi_m$ with fixed $J_{1y}=0.4$ for the AFM--VBS transition and with fixed $J_{1y}=0.85$ for the AFM--QSL and QSL--VBS transitions as examples. Generally, for the AFM--VBS transition, subleading corrections are needed for a good data collapse. (However, the case $J_{1y}=0.4$ does not need such a correction.) For the AFM--QSL and QSL--VBS transitions, we find that subleading terms are always unnecessary. 

The extracted critical exponents for different $J_{1y}$ cases are summarized in  Table ~\ref{tab:criticalexponents}. For the AFM--VBS transition, the critical exponents for $\ave{M^2_0}$ and $\ave{D^2_x}$ indicate that $z+\eta_{s}^*=z+\eta_{d}^*\sim 1.36$. For the AFM--QSL and QSL--VBS transitions, the exponents are clearly different from those of the AFM--VBS transition. Roughly speaking, $z+\eta_{s1}^*\sim 1.2$ and $z+\eta_{d1}^*\sim 1.8$ for the AFM--QSL transition and $z+\eta_{s2}^*=z+\eta_{d2}^* \sim 1.5$ for the QSL--VBS transition. For $J_{1y}=0.6$ very close to the tricritical point, the exponents are slightly different, which might be due to finite-size effects. 
We note that $z+\eta_{s}^*=z+\eta_{d}^*$ and $z+\eta_{s2}^*=z+\eta_{d2}^*$ also consistent with the emergent O(4) symmetry. 
The obtained critical exponents including $\nu\approx 1.0$ strongly support new universality classes for the AFM--QSL and QSL--VBS transitions, which are different from the class for the DQCP.
   \begin{table}[htbp]
   \centering
 \caption { Critical exponents of the $J_{1x}$-$J_{1y}$-$J_2$ model  at the AFM--QSL and QSL--VBS transition points or at the AFM--VBS transition  using fixed $J_{1y}$. Errors in exponents are from fitting. Errors in the critical point $J_c$ are estimated from crossing points of physical quantities or  fitting correlation lengths. Values of $\nu$ for QSL-related transitions are averages of several values obtained from fitting physical quantities; see the Supplemental Information. Spin and dimer exponents $z+\eta^*$ are obtained through data collapse using  the listed values of $\nu$ and $J_c$.}
	\begin{tabular*}{\hsize}{@{}@{\extracolsep{\fill}}lccccc@{}}
		\hline\hline
	   model &type  &   $z+\eta_s^*$ & $z+\eta_d^*$    & $\nu$ & $J_c$  \\ \hline
 		
         $J_{1y}=0.2$ &AFM--VBS & 1.36(1) & 1.36(2) & 0.84(5)  & 0.071(2)\\
         $J_{1y}=0.4$ &AFM--VBS & 1.36(4) & 1.38(3) & 0.85(6) & 0.171(2) \\
         $J_{1y}=0.55$ &AFM--VBS & 1.35(1) & 1.34(2) & 0.85(5) & 0.255(2)\\  \\
          $J_{1y}=0.60$ & AFM--QSL  & 1.36(1)   & 1.55(1)  & 0.97(4) & 0.273(4) \\
	   	$J_{1y}=0.60$ &QSL--VBS &1.44(2)   & 1.45(1)  & 0.97(4) & 0.283(1) \\ \\
	     $J_{1y}=0.65$ & AFM--QSL  & 1.23(2)   & 1.70(1)  & 1.00(5) & 0.285(2) \\
	   	$J_{1y}=0.65$ &QSL--VBS &1.49(1)   & 1.49(1)  & 1.00(5) & 0.309(1) \\ \\
      $J_{1y}=0.75$ & AFM--QSL  & 1.23(1)   & 1.74(1)  & 1.01(4) & 0.327(2)   \\
	   	$J_{1y}=0.75$ &QSL--VBS  &1.47(1)   & 1.47(1)  & 1.01(4) &0.360(1)  \\ \\
        $J_{1y}=0.85$ & AFM--QSL  & 1.18(1)   & 1.85(2)  & 1.00(4) &0.352(5)  \\
	   	$J_{1y}=0.85$ &QSL--VBS  &1.50(1)   & 1.50(2)  & 1.00(4) & 0.402(3) \\ \\
	   	 $J_{1y}=0.95$ & AFM--QSL  & 1.21(2)   & 1.88(2)  & 1.05(5) & 0.390(2) \\
 		$J_{1y}=0.95$ &QSL--VBS  &1.52(1)   & 1.52(2)  & 1.05(5) & 0.453(2) \\  \\
 		$J_{1y}=0.98$ & AFM--QSL  & 1.27(1)   & 1.86(2)  & 1.00(4) & 0.410(5) \\
 		$J_{1y}=0.98$ &QSL--VBS  &1.52(2)   & 1.52(2)  & 1.00(4) & 0.465(3) \\  
 		\hline\hline
	\end{tabular*}
\label{tab:criticalexponents}	
\end{table}

\bigskip
 \noindent\textbf{Discussion}\\
\noindent
In summary, we study the $J_{1x}$-$J_{1y}$-$J_2$ model using the state-of-the-art tensor network method. In the strong-anisotropy region, we identify a continuous phase transition line between the AFM and columnar VBS phase, where emergent O(4) symmetry appears. With weakening anisotropy, the AFM--VBS transition line terminates at a tricritical point, from which a gapless QSL emerges  between the AFM and VBS phases. Most surprisingly, we find that the emergent O(4) symmetry persists on the QSL--VBS phase boundary. We stress that the discovered QSL phase cannot be a finite-size effect for the following reasons. First, a peculiar point located at $(J_{1y},J_2)\simeq (0.6,0.3)$ was suggested by a previous study using the coupled cluster (CC) method~\cite{Bishop2008}, and this point is close to our estimated tricritical point at $(J_{1y},J_2)\simeq (0.58,0.27)$. Second, recent studies have consistently supported the appearance of a QSL phase in the $J_1$-$J_2$ and $J_1$-$J_2$-$J_3$ models ~\cite{gong2014,wang2017,Ferrari2020,yusuke2021,liuQSL,liuj1j2j3}. Third, the emergent O(4) symmetry on the QSL--VBS boundary does not appear on the AFM--QSL boundary, which helps us clearly identify the QSL region. 
We note that in the $J_1$-$J_2$-$J_3$ model with $C_4$ lattice symmetry, the emergent SO(5) symmetry seems not to appear on the QSL--VBS phase boundary~\cite{liuj1j2j3}, and precise numerical calculation of the correlation length exponent gives $\nu\sim 0.45$ for the SO(5) deconfined transition ~\cite{loopmodel1,JQ2016,JQ2020}, which is inconsistent with the conformal bootstrap constraint $\nu>0.51$~\cite{conformalbootstrap}, suggesting a weakly first-order transition in the thermodynamic limit. 
In the $J_{1x}$-$J_{1y}$-$J_2$ model, since the QSL--VBS phase transition is unlikely to be weakly first order, we believe that such an emergent O(4) symmetry should survive in the thermodynamic limit. 

Constructing a quantum field theory description for both the QSL and DQCP  with emergent O(4) symmetry is challenging in that it involves three different types of unconventional phase transition and a tricritical point. There have been theoretical attempts to understand the phase diagram for the $J_1$-$J_2$-$J_3$ model with $C_4$ lattice symmetry~\cite{shackleton2021,shackleton2022anisotropic}, but the theoretical predictions have contradicted the numerical results~\cite{shackleton2021,liuj1j2j3}. For the $J_{1x}$-$J_{1y}$-$J_2$ model, the emergent O(4) symmetry at the AFM--VBS and QSL--VBS transitions provides a strong constraint for future theoretical studies. Specifically, in the strong-anisotropy region, the $J_{1x}$-$J_{1y}$-$J_2$ model comprises weakly coupled spin-1/2 chains~\cite{Tsvelik2003,balents2004}. This well-understood model provides a starting point for understanding the emergent O(4) symmetry. Remarkably, an enlarged symmetry formed by spin and dimer order parameters has been indicated in a chain-mean-field study, although not rigorously established~\cite{balents2004}. Experimentally, the $J_{1x}$-$J_{1y}$-$J_2$ model can be realized for cold atoms by coupling one-dimensional spin-1/2 chains~\cite{blatt2012quantum,bloch2012quantum}, and the verdazyl-based salt [$m$-MePy-V-($p$-F)$_2$]SbF$_6$ is a real material having application potential~\cite{anisomaterial2021}.

\bigskip
 \noindent\textbf{Methods}\\
\noindent
\noindent\textbf{Tensor Network Method.} The tensor network state, specifically, projected entangled pair state (PEPS), offers a powerful description for entangled quantum many-body states~\cite{verstraete2008}, and has been extensively employed to characterize various types of states, including exotic topologically ordered phases. As an extension of the one-dimensional density matrix renormalization group (DMRG) method to higher dimensions, PEPS can efficiently capture the entanglement structure of 2D systems with systematically improvable precision controlled by the tensor bond dimension $D$, and provides an excellent approach for the simulation of frustrated magnets where Quantum Monte Carlo (QMC) fails. We use the method of finite PEPS in the scheme of variational Monte Carlo, detailed in Ref.~\cite{liu2017,liufinitePEPS}. Such an approach has been demonstrated as a powerful way for finite size calculations, through massive comparisons with available QMC, iPEPS, and density matrix renormalization group (DMRG) results on various physical models~\cite{liu2017,liu2018,liufinitePEPS,liuQSL,liuj1j2j3}. Tensor bond dimension $D=8$ of PEPS is adopted for the simulations, which works excellently with high precision for the presented system sizes on other unfrustrated and frustrated systems~\cite{liufinitePEPS,liuQSL,liuj1j2j3}. This work took approximately 10 million CPU hours, as it is necessary to sweep a two-dimensional space of tuning parameters $(J_{1y},J_2)$ for different systems $L\times L$ with $L=6-20$.  

\begin{figure}[htbp]
 \centering
 \includegraphics[width=3.4in]{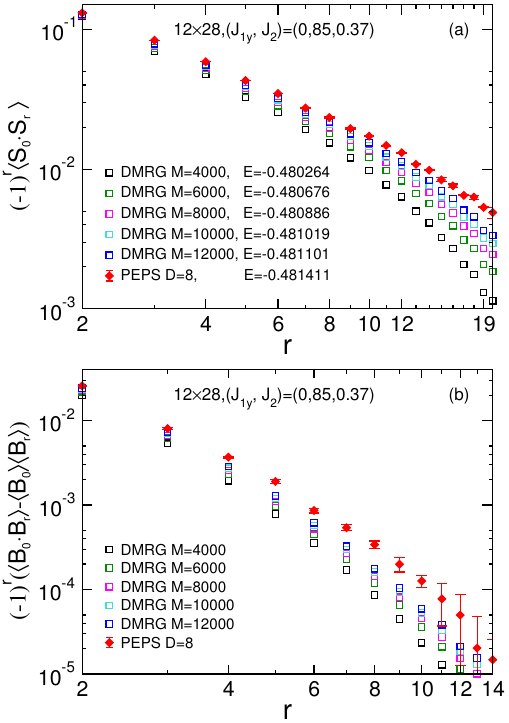}
 \caption{Comparison of spin and dimer correlations between PEPS and DMRG on a $12\times 28$ lattice at $(J_{1y},J_2)=(0.85,0.37)$. Energy persite is presented in the legend.}
 \label{fig:DMRG_PEPS}
 \end{figure}

To further check the accuracy of the finite PEPS method with a bond dimension $D=8$, the obtained results are compared with DMRG results on a long strip with $L_y=12$ and $L_x=28$, which is almost at the width $L_y$ limit of the DMRG method. The point $(J_{1y},J_2)=(0.85,0.37)$  within the region of the QSL phase is chosen as the reference,  which is believed to be very difficult for accurate simulation. The energies for different bond dimensions $M$ of the DMRG are listed in Fig.~\ref{fig:DMRG_PEPS} (a), for comparison with the energy of the PEPS obtained using $D=8$. The DMRG incorporating SU(2) spin rotation symmetry is used, such that the largest bond dimension $M=12,000$ is equivalent to 48,000 U(1) states. All the comparisons of the energy, spin, and dimer correlations suggest that the PEPS with $D=8$ provides excellent results in the critical phase. It is thus reasonable that $D=8$ gives very good results for other cases with similar sizes. In fact, previous extensive comparisons of DMRG and PEPS methods applied to the spin-1/2 square lattice $J_1$-$J_2$ model and $J_1$-$J_2$-$J_3$ model of Heisenberg antiferromagnets, as well as a comparison of results for PEPS $D=4$ to 10, explicitly demonstrated that setting $D=8$ enables good convergence of the results for system sizes up to $20\times 28$ in highly frustrated regions and for system sizes up to $32\times 32$ in an unfrustrated Heisenberg model~\cite{liufinitePEPS,liuQSL,liuj1j2j3}. 

\bigskip
 \noindent\textbf{Data availablity}\\
\noindent
The data that support the findings of this study are available
from the corresponding authors upon request.

\bigskip
 \noindent\textbf{Acknowledgments} \\
 \noindent
 We thank Subir Sachdev, Yin-Chen He, Chong Wang and Cenke Xu for helpful discussions. We also thank Didier Poilblanc for related work. This work was supported by the NSFC/RGC Joint Research Scheme No. N-CUHK427/18 of the Hong Kong Research Grants Council and No. 11861161001 of the National Natural Science Foundation of China.
WQC was supported by the National Key R\&D Program of China (Grants No. 2022YFA1403700), Science, Technology and Innovation Commission of Shenzhen Municipality (under grant ZDSYS20190902092905285), Guangdong Basic and Applied Basic Research Foundation (under grant 2020B1515120100), and Center for Computational Science and Engineering at Southern University of Science and Technology.
S.S.G. was supported by the National Natural Science Foundation of China Grants No. 11874078 and No. 11834014.  

\bigskip
\noindent\textbf{Author contributions} \\
\noindent
 Wenyuan Liu carried out the PEPS simulations; Shoushu Gong carried out the DMRG calculations. Weiqiang Chen and Zhengcheng Gu supervised the project. Wenyuan Liu and Zhengcheng Gu wrote the
manuscript with input from Shoushu Gong and Weiqiang Chen. All the authors participated in the discussion.

\bigskip
 \noindent\textbf{Competing interests}\\
  \noindent
 The authors declare no competing interests.

\clearpage

\appendix

\section{1. VBS--stripe phase transition}
\subsection{A. $J_{1y}=0.95$}
\begin{figure}[htbp]
 \centering
 \includegraphics[width=3.4in]{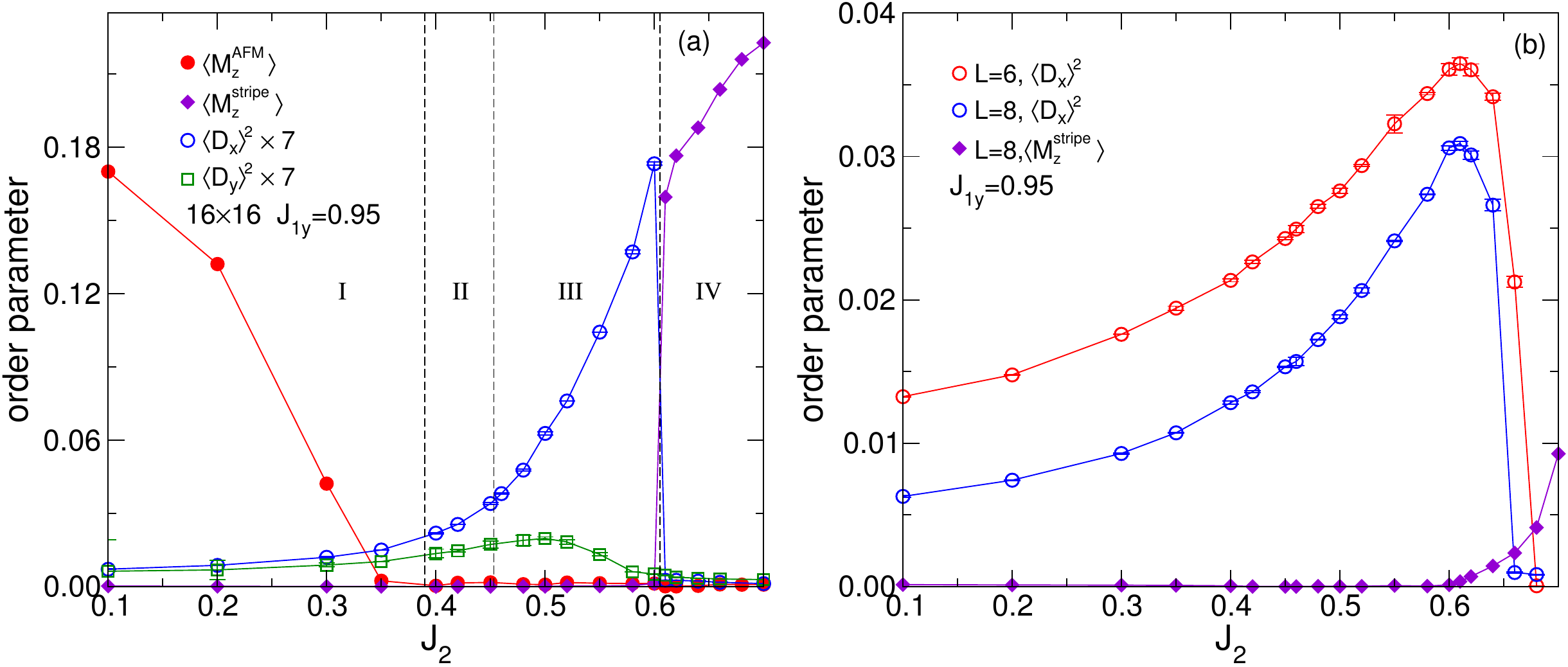}
 \caption{(a) $J_2$ dependence of local order parameters on a $16\times 16$ lattice at $J_{1y}=0.95$, including N\'eel AFM order parameters $\langle M^{\rm AFM}_{z}\rangle $, boundary-induced VBS order parameters $\langle D_x\rangle^2$ and $\langle D_y\rangle^2$ (both magnified by a factor of 7 for clarity), and collinear AFM order parameters $\langle M^{\rm stripe}_{z}\rangle $ for the stripe phase. Black dashed lines separate four phases: (I) AFM,  (II) QSL, (III) columnar VBS, and (IV) stripe. (b) $J_2$ dependence of local order parameters on small $6\times 6$ and $8\times 8$ lattices.  }
 \label{fig:VBS_stripe_J1y_095}
 \end{figure}
 
 We consider the phase diagram with respect to $J_2$  at fixed $J_{1y}=0.95$. Taking the ground states on a $16\times 16$ lattice as an example, we show how the local order parameters change. In our calculations, we always sample in the $S^{z}_{\rm tot}=\sum_{\bf i}S^z_{\bf i}$ subspace, which is equivalent to imposing U(1) symmetry on the wave function, and we thus need only to consider the $z-$component of local magnetic order parameters.    
 The magnetic properties of the AFM and stripe phases can be determined using the local  N\'eel AFM order parameters,
 \begin{equation}
 M^{\rm afm}_z=\frac{1}{L^2}\sum_{\bf i}(-1)^{i_x+i_y}S^z_{\bf i},
 \end{equation}
 and collinear AFM order parameters,
  \begin{equation}
 M^{\rm stripe}_z=\frac{1}{L^2}\sum_{\bf i}(-1)^{i_y}S^z_{\bf i}.
 \end{equation}
 Note that $ M^{\rm afm}_z\equiv M_z$ in our notation.  The possible VBS pattern is reflected by boundary-induced order parameters $\langle D_{\alpha}\rangle^2, $where $\alpha=x,y$.
 
  Figure~\ref{fig:VBS_stripe_J1y_095} presents the local order parameters in the whole $J_2$ region of interest, which includes four phases: AFM, QSL, VBS, and stripe phases. The vertical dashed lines for AFM--QSL and QSL3--VBS transitions denote the phase boundaries obtained in the thermodynamic limit, and the other vertical dashed line for the VBS--stripe transition is the phase boundary from $16\times 16$.  Figure~\ref{fig:VBS_stripe_J1y_095}(a) shows that $\langle  M^{\rm afm}_z \rangle$ and $\langle  M^{\rm stripe}_z \rangle$   have  large values in the AFM and stripe phases, respectively, whereas they are  almost zero in other phases. Theoretically, these values should be zero in each phase because in finite systems, the exact ground state  should be a singlet. In small systems like the $6\times 6$ lattice, the local magnetic order indeed is almost zero.  The obvious nonzero values in magnetic phases  in large systems are in fact a reflection of the spontaneous symmetry breaking in the thermodynamic limit. Phenomena of spontaneous symmetry breaking are also observed in finite size calculations and have already been fully discussed for the DMRG method~\cite{white2021}.  The almost zero  local magnetic order parameters in QSL and VBS phases indicate the recovery of spin rotation symmetry, and we indeed find that $\langle M^2_{\alpha} \rangle=\frac{1}{3}\langle M^2_0 \rangle$ ($\alpha=x,y,z$).  
 
The boundary-induced VBS order parameters $\langle D_{x} \rangle^2$ and  $\langle D_{y} \rangle^2$  are such that  $\langle D_{x} \rangle^2$  is larger than $\langle D_{y} \rangle^2$, especially in the VBS phase, which is a clear signature of anisotropy. Of course, $\langle D_{y}\rangle^2$ always has a zero extrapolated value for $L\rightarrow \infty$.  Interestingly, different from  $\langle D_{x} \rangle^2$, in the VBS phase, $\langle D_{y} \rangle^2$ first increases and then decreases at some $J_2$ (here, the $\langle D_{y} \rangle^2$  peak is around $J_2=0.5$), and  $\langle D^2_{y} \rangle$ has the same behavior. Note that $\langle D_{y} \rangle^2$  and  $\langle D^2_{y} \rangle$ are scaled to zero in the thermodynamic limit.

The VBS--stripe phase transition is a typical first-order transition. The variations in local order parameters on $6\times 6$, $8\times 8$, and $16\times 16$ lattices are presented in Fig.~\ref{fig:VBS_stripe_J1y_095}. In all the cases, an increase in $J_2$ results in a sharp change in $\langle D_{x} \rangle^2$ near $J_2=0.60$, with $\langle D_{x} \rangle^2$ falling to zero. On the $6\times 6$ lattice, $\langle  M^{\rm stripe}_z \rangle$ is almost zero for all presented $J_2$, whereas on larger systems, $\langle  M^{\rm stripe}_z \rangle$ has nonzero values in the stripe phase, indicating that the spin rotation symmetry is broken. The transition point $J_{c3}(L)$ at different sizes $L$ shifts with increasing $L$, as seen for the previous $J_1$-$J_2$  model~\cite{liuQSL}. The 2D limit transition point $J_{c3}$ is easily evaluated at $J_{c3}\simeq0.573$ by using the methods from Ref.\cite{liuQSL}.
\begin{figure}[htbp]
 \centering
 \includegraphics[width=3.4in]{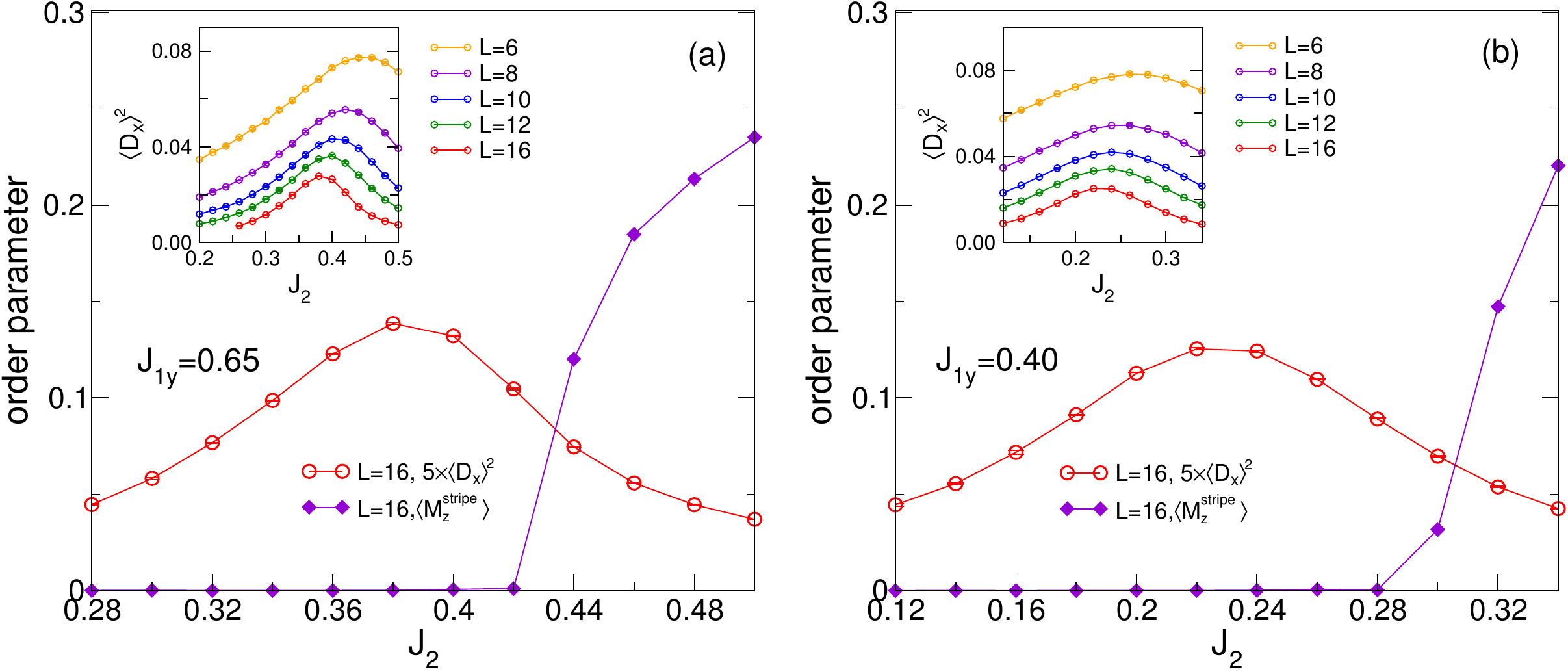}
 \caption{$J_2$ dependence of  $\ave{D_x}^2$ (times a factor of 5) and $\langle M^{\rm stripe}_{z}\rangle $ on a $16\times 16$ lattice for $J_{1y}=0.65$ (a) and $J_{1y}=0.4$ (b). Insets show $\ave{D_x}^2$ for $L=6-16$. }
 \label{fig:VBS_stripe_J1y_065_040}
 \end{figure}
 
\subsection{B. $J_{1y}=0.65$ and $J_{1y}=0.40$}
We consider large anisotropy regions corresponding to say $J_{1y}$=0.65 and 0.4. It is seen that the VBS--stripe transition is still of first order, as clearly signaled by the behavior of $\langle  M^{\rm stripe}_z \rangle$. For $\ave{D_x}^2$, the peak gradually becomes narrower as the system size increases, which is consistent with a first-order transition. Compared with the case for $J_{1y}=0.95$, the peaks of $\ave{D_x}^2$  for $J_{1y}$=0.65 and 0.4 on the $16\times 16$ lattice are much broader. This indicates that the transition is not as strong as that for $J_{1y}=0.95$. A comparison of the broadness of the $\ave{D_x}^2$ peak on the $16\times 16$ lattice for $J_{1y}=0.95$, 0.65, and 0.4 suggests that with $J_{1y}$ decreasing, the transition becomes gradually weaker. 
\subsection{C. Comparison with early studies}

Our overall phase diagram combines previous seemingly conflicting  results in a consistent way. Analytical and exact diagonalization (ED) studies have suggested a VBS phase between the AFM and stripe phases for all $J_{1y}\leq 1$~\cite{balents2004,Sindzingre2004}. In fact, the analytical results hold only for small $J_{1y}$ and the ED results are limited to small system sizes. Our results agree well with the results of these studies at small $J_{1y}$. Additionally, a CC analysis has suggested a direct continuous transition between AFM and stripe phases below a particular point at $(J_{1y},J_2)\simeq (0.6,0.3)$, and above the  point where the AFM and stripe phases are separated by a nonmagnetic phase~\cite{Bishop2008}. Our tensor network results show that at small $J_{1y}$ there exists a VBS phase between the AFM and stripe phases. The CC results in this region  contradict our results, as well as the aforementioned analytical and ED results~\cite{balents2004,Sindzingre2004}. The suggested continuous AFM--stripe phase transition in the CC results violates both Landau and existing DQCP paradigms. Note that we find that the VBS--stripe phase transition is weak at small $J_{1y}$. The discovered VBS phase between the AFM and stripe phases and the weakness of the VBS--stripe transition adequately explain why a continuous AFM--stripe phase transition is observed in the CC study. Hence, our results reconcile the CC results with the results of other analyses, and place all the phase transitions in the Landau paradigm and DQCP paradigm. Finally, the nonmagnetic phase suggested by the CC results in fact contains a QSL phase and a VBS phase according to our results. This finding connects the results of the $J_1$-$J_2$ model and small-$J_{1y}$ results through an AFM--VBS transition via a tricritical point.
 
\section{2. Gapless QSL region }
\subsection{A. Crossing points}
\begin{figure*}[htbp]
 \centering
 \includegraphics[width=6.8in]{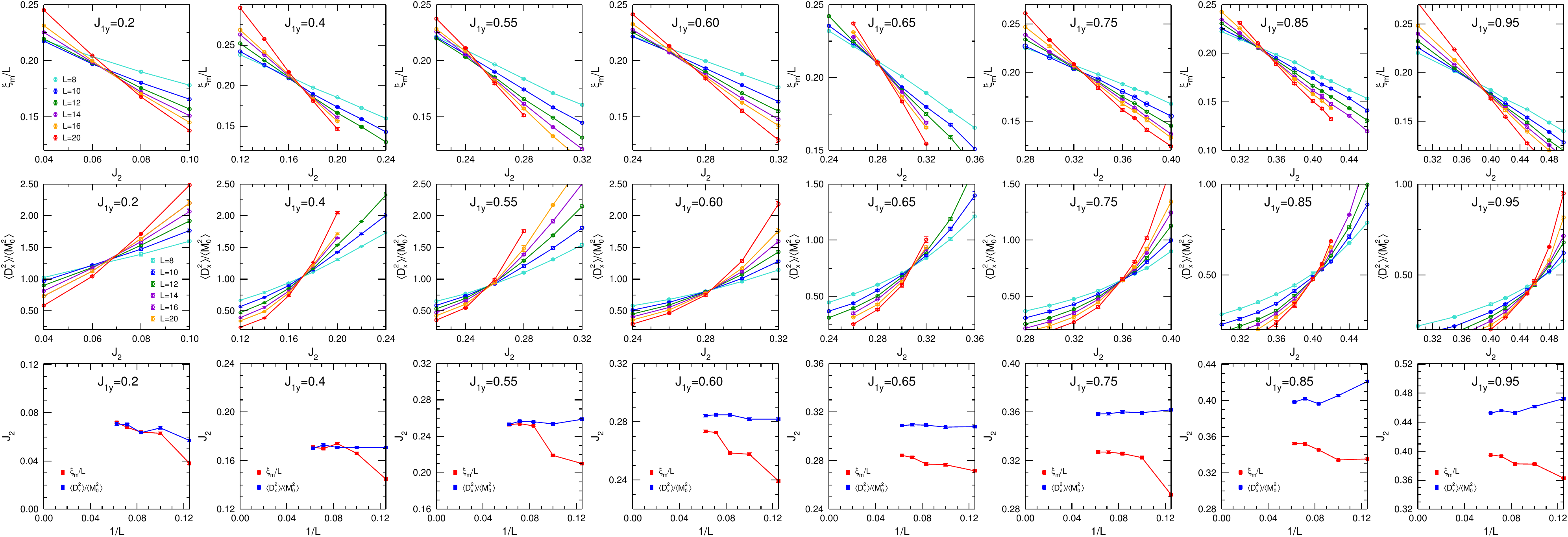}
 \caption{Crossing points for dimensionless quantities $\xi_m/L$ (top row) and  $\langle D^2_x \rangle / \langle M^2_0\rangle$ (middle row) at different fixed $J_{1y}$. The bottom row shows the  
 scaling of crossing points between the sizes of $L$ and $L+2$ (except at $L=16$, where  crossing points between $L=16$ and $L=20$ are shown). }
 \label{fig:crossing}
 \end{figure*}
We show the crossings from  $\xi_m /L$ and  $\langle D^2_x \rangle / \langle M^2_0\rangle$ and analyze their finite effects by plotting  crossing points from $(L,L+2)$ for $L\leq 14$ and $(L,L+4)$ for $L=16$ versus $1/L$  in Fig.~\ref{fig:crossing}.
Generally, the crossing $J_2$ values of   $\xi_m /L$  give the AFM--VBS transition point $J_2=J_{c1}$, as expected. This result is consistent with the  finite-size extrapolation results of AFM order parameters for small $J_{1y}=0.2,$ 0.4, and 0.55. The  quantities of  $\xi_m /L$ and  $\langle D^2_x \rangle / \langle M^2_0\rangle$ give the same crossing $J_2$ value for each case, which is consistent with a direct AFM--VBS transition with emergent O(4) symmetry. For larger $J_{1y}\geq 0.65$, it is seen  that in this context the crossing $J_2$ values of $\langle D^2_x \rangle / \langle M^2_0\rangle$ are different from those of $\xi_m/L$; see Fig.~\ref{fig:crossing}. Considering the coincidence of crossing $J_2$ values  between   $\xi_m /L$ and   $\langle D^2_x \rangle / \langle M^2_0\rangle$ at AFM--VBS transitions, the discrepancies for larger $J_{1y}$ provide strong evidence for the existence of a different phase, namely the QSL phase.

A  feature of great interest is that the crossing $J_2$ values of $\langle D^2_x \rangle / \langle M^2_0\rangle$   are almost the same as  the QSL--VBS transition points $J_2=J_{c2}$ obtained by the finite-size scaling of  VBS order parameters. This motivates us to evaluate the crossing $J_2$ values  as precisely as possible. We note that the crossing values with respect to $1/L$ cannot be fitted well by a simple polynomial function, possible due to the imperfect optimization of wave functions. Nevertheless, the crossing points from  $\langle D^2_x \rangle / \langle M^2_0\rangle$ have very small finite-size effects, which has also been observed in other studies~\cite{loopmodel2,sreejith2019}. This enables us to evaluate the thermodynamic limit value  by simply averaging the large-size values. For $\xi_m/L$, the crossing $J_2$ values for large system sizes can be estimated as the thermodynamic limit transition point $J_{c1}$.  We also use the collective fitting of $J_{c1}$ and correlation length $\nu$ for data collapse to take into account the finite-size effects and obtain consistent results. Furthermore, we compute more $J_2$ points to reduce the uncertainty in finite size scaling for locating the VBS phase boundary, as seen in the insets of 
Fig.2(d) in the main text, as well as Fig.~\ref{fig:DataCollapse_J1y_075}(b), and Fig.~\ref{fig:DataCollapse_J1y_095}(b). 
 The values are summarized in Table. I 
in the main text. We see that for several different cases with fixed $J_{1y}=0.60$, 0.65, 0.75, 0.85, 0.95, and 0.98, within our resolution, the ratio $\langle D^2_x \rangle / \langle M^2_0\rangle$ and finite size scaling of $\langle D^2_x \rangle$ give the same results. The quantity $\langle D^2_x \rangle / \langle M^2_0\rangle$ is only related to the emergent symmetry, which indicates that the QSL--VBS phase transition points have emergent O(4) symmetry.
 \subsection{B. Finite size scaling of order parameters}
 
\begin{figure}[htbp]
 \centering
 \includegraphics[width=3.4in]{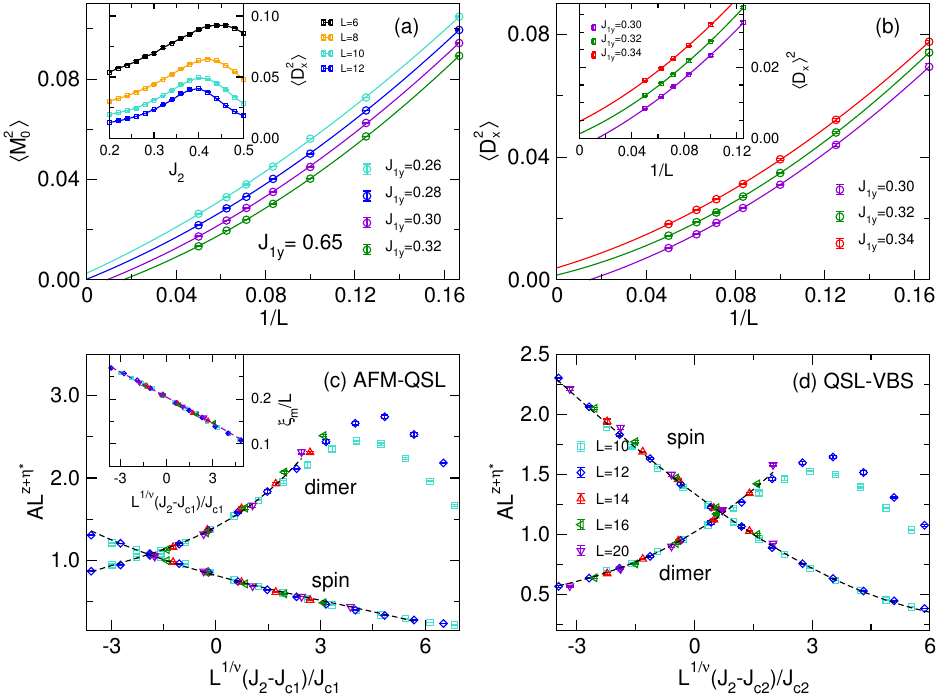}
 \caption{Order parameters of the $J_{1x}$-$J_{1y}$-$J_2$ model at fixed $J_{1y}=0.65$. (a) Finite size scaling for AFM order parameters (main panel) and the variance in VBS order parameters $\langle D^2_x\rangle $ with respect to $J_2$ (inset). (b)  Finite-size scaling for VBS order parameters $\langle D^2_x\rangle $ (main panel) and $\langle D_x\rangle^2$ (inset).  Second-order polynomial fits are used. (c) Data collapse of AFM and VBS order parameters and the correlation length $\xi_m$ at AFM--QSL transition point $J_{c1}=0.285$, with $\nu=1.00$,  $z+\eta_{s1}^*=1.23$, and $z+\eta_{d1}^*=1.70$. (d) Data collapse of AFM and VBS order parameters at QSL--VBS transition point $J_{c2}=0.309$, with $\nu=1.00$ and  $z+\eta_{s2}^*=z+\eta_{d2}^*=1.49$. Black dashed curves in (c) and (d) are  second-order curves  with corresponding critical exponents. }
 \label{fig:DataCollapse_J1y_065}
 \end{figure}
 
 We consider the $J_{1x}$-$J_{1y}$-$J_2$ model at fixed $J_{1y}=0.65$. Figure~\ref{fig:DataCollapse_J1y_065}(a) shows that the AFM order parameter vanishes around $J_2\simeq 0.28$; this is consistent with the behavior of $\xi_m/L$, which gives $J_2\simeq 0.285$. Meanwhile, the finite-size scaling of VBS order parameters $\langle D^2_x\rangle$ shows that the VBS order begins to appear between $J_2=0.30$ and $J_2=0.32$, which is confirmed by the boundary-induced order parameters $\langle D_x\rangle^2$, as presented in  Fig.~\ref{fig:DataCollapse_J1y_065}(b). The ratio $\langle D^2_x\rangle/ \langle M^2_0\rangle$ gives a crossing value $J_2\simeq 0.309$,  which is well located within the region $[0.30,0.32]$.   Additionally, we show the VBS order parameters $\ave{D^2_x}$ w.r.t. $J_2$ on different systems with $L=6$,8,10, and 12  in the inset of Fig.~\ref{fig:DataCollapse_J1y_065}(a). The order parameters have peaks for all systems, indicating the VBS--stripe transition point. The corresponding boundary-induced dimerizations $\ave{D_x}^2$ have already been presented in Fig.~\ref{fig:VBS_stripe_J1y_065_040}.  When we scale the VBS order parameters for data collapse, the values near the peaks do not collapse well, as shown in  Fig.~\ref{fig:DataCollapse_J1y_065}(c) and (d), perhaps because they are far from the critical region.
 \begin{figure}[htbp]
 \centering
 \includegraphics[width=3.4in]{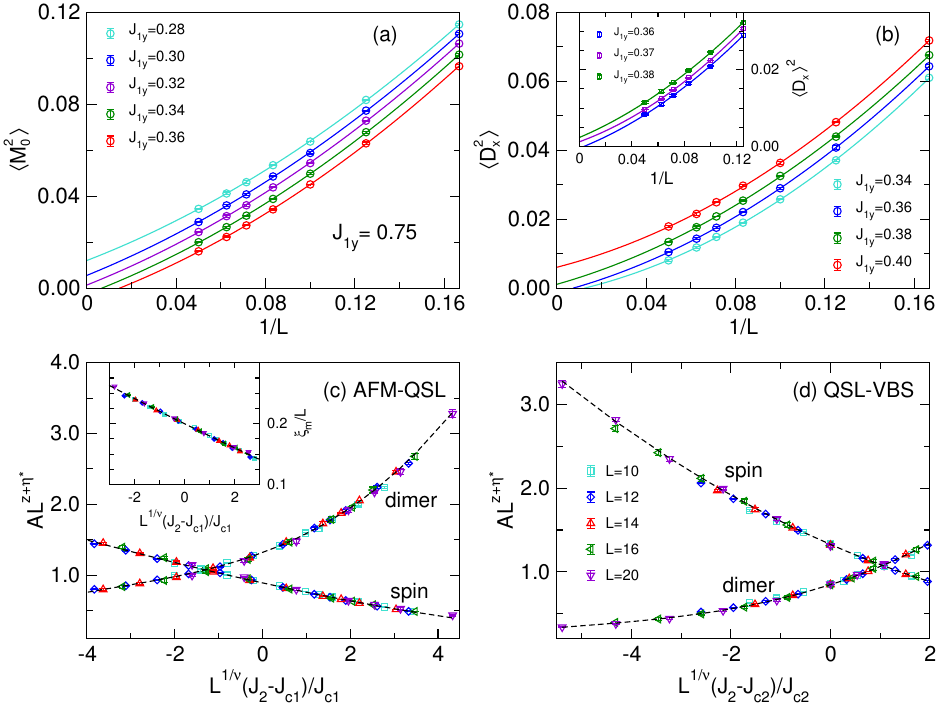}
 \caption{Scaling of spin and dimer order parameters of the $J_{1x}$-$J_{1y}$-$J_2$ model at fixed $J_{1y}=0.75$ for AFM--QSL (a,c)  and QSL--VBS (b,d) transitions. Panels(a) and (b) show the finite-size scaling for AFM and VBS order parameters, $\langle M^2_0\rangle $ and $\langle D^2_x\rangle $. The inset of (b) shows the boundary-induced VBS order parameters $\langle D_x\rangle^2$. Second-order polynomial fits are used. (c) Data collapse of AFM and VBS order parameters and the correlation length $\xi_m$ at the AFM--QSL transition point $J_{c1}=0.327$, with $\nu=1.01$,  $z+\eta_{s1}^*=1.23$, and $z+\eta_{d1}^*=1.74$. (d) Data collapse of AFM and VBS order parameters at the QSL--VBS transition point $J_{c2}=0.360$, with $\nu=1.01$ and  $z+\eta_{s2}^*=z+\eta_{d2}^*=1.47$. Black dashed curves in (c) and (d) are  second-order curves  with corresponding critical exponents. }
 \label{fig:DataCollapse_J1y_075}
 \end{figure}

 \begin{figure}[htbp]
 \centering
 \includegraphics[width=3.4in]{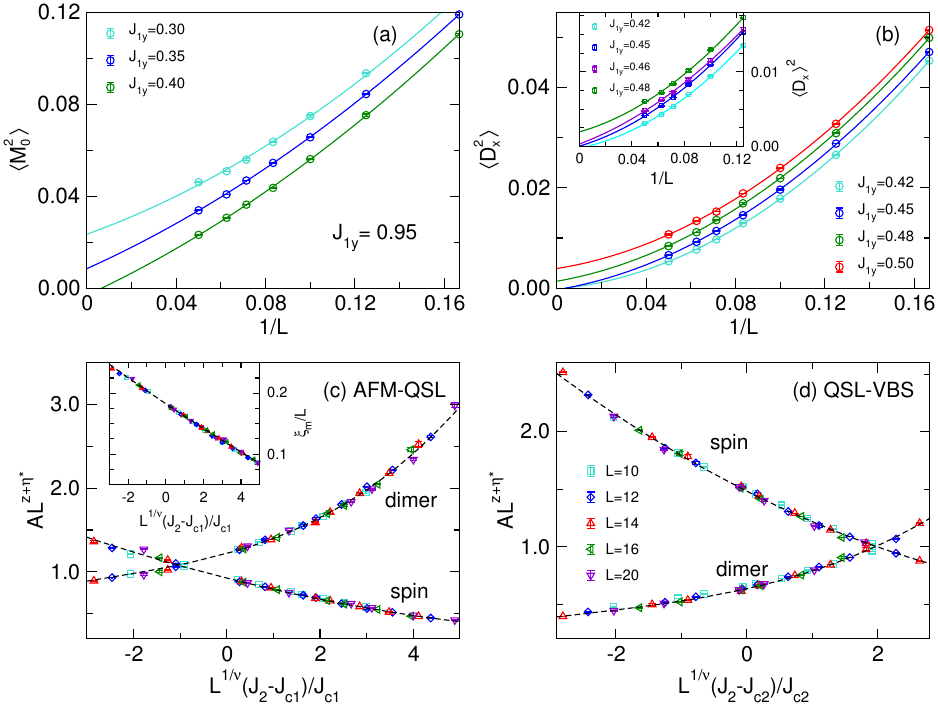}
 \caption{Scaling of spin and dimer order parameters of the $J_{1x}$-$J_{1y}$-$J_2$ model at fixed $J_{1y}=0.95$ for AFM--QSL (a,c)  and QSL--VBS (b,d) transitions. Panels (a) and (b) show the finite-size scaling for AFM and VBS order parameters, $\langle M^2_0\rangle $ and $\langle D^2_x\rangle $. The inset of (b) shows boundary-induced VBS order parameters $\langle D_x\rangle^2$. Second-order polynomial fits are used. (c) Data collapse of AFM and VBS order parameters and the correlation length $\xi_m$ at the AFM--QSL transition point $J_{c1}=0.39$, with $\nu=1.05$,  $z+\eta_{s1}^*=1.21$, and $z+\eta_{d1}^*=1.88$. (d) Data collapse of AFM and VBS order parameters at the QSL--VBS transition point $J_{c2}=0.453$, with $\nu=1.05$ and  $z+\eta_{s2}^*=z+\eta_{d2}^*=1.52$. Black dashed curves in (c) and (d) are  second-order curves  with corresponding critical exponents. }
 \label{fig:DataCollapse_J1y_095}
 \end{figure}

 Similarly, we make computations for other fixed $J_{1y}=0.75$, 0.85, and 0.95. The results for $J_{1y}=0.75$ and $J_{1y}=0.95$ are presented in Fig.~\ref{fig:DataCollapse_J1y_075}(a--d) and Fig.~\ref{fig:DataCollapse_J1y_095}(a--d), respectively.   Note that the simulated systems are under open boundary conditions. The dimer structure factor is not well defined in this situation and hence one cannot obtain the corresponding dimer correlation length to determine the VBS boundaries~\cite{zhao2020,liuQSL}. For the QSL--VBS transition, to compare with the crossing $J_2$ value of the ratio $\langle D^2_x\rangle/ \langle M^2_0\rangle$,  we try our best to reduce the uncertainty in the transition point by computing more  $J_2$ points, as well as using different fitting functions and different system sizes for the finite size scaling of $\langle D^2_x\rangle$ and $\langle D_x\rangle^2$.  As stated in the main text, the results suggest that the crossing $J_2$ value for $L\rightarrow \infty$ from  $\langle D^2_x\rangle/ \langle M^2_0\rangle$ is the same as that from finite size scaling within our resolution.

 \begin{figure}[htbp]
 \centering
 \includegraphics[width=3.4in]{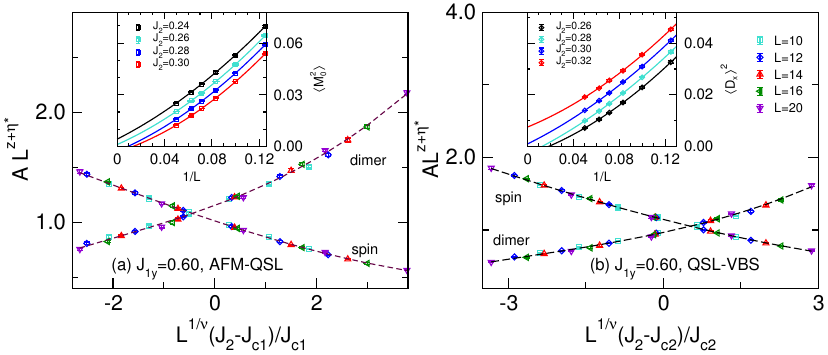}
 \caption{Scaling  of spin and dimer order parameters of the $J_{1x}$-$J_{1y}$-$J_2$ model at fixed $J_{1y}=0.6$  for the AFM--QSL (a) and QSL--VBS (b) transitions. The main panels show the data collapse and the insets show the finite size scaling of  $\langle M^2_0\rangle$ and  $\langle D^2_x\rangle$ with second-order fits. At the AFM--QSL transition (a), $\nu=0.97$, $J_{c1}=0.273$ , $z+\eta_s^*=1.36$, and $z+\eta_d^*=1.55$  are used for the quantities  $\langle M^2_0\rangle$ and  $\langle D^2_x\rangle$. At the QSL--VBS transition (b), $\nu=0.97$, $J_{c2}=0.283$ , and $z+\eta_s^*=z+\eta_d^*=1.45$.  Black dashed lines are fitted curves  with corresponding critical exponents. }
 \label{fig:DataCollapse_J1y_060}
 \end{figure}
\begin{figure}[htbp]
 \centering
 \includegraphics[width=3.4in]{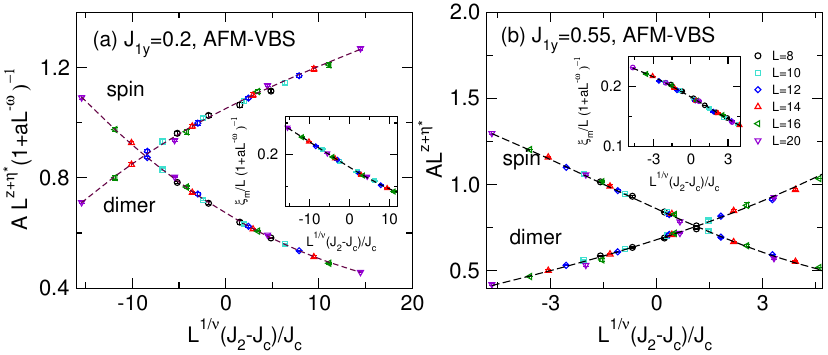}
 \caption{Scaling analysis of spin and dimer order parameters of the  $J_{1x}$-$J_{1y}$-$J_2$ model at fixed $J_{1y}=0.2$ (a) and $J_{1y}=0.55$ (b) for the AFM--VBS transition. Insets show respective scaling of the correlation lengths. At $J_{1y}=0.2$, using $\nu=0.84$, $J_c=0.071$, $z+\eta_s^*=z+\eta_d^*=1.36$, and $\omega=2$. The prefactors $a$ are 4.5, 8, and 0 for the quantities $\xi_m$, $\langle M^2_0\rangle$, and $\langle D^2_x\rangle$, respectively. At $J_{1y}=0.55$, using $\nu=0.85$, $J_c=0.255$, $z+\eta_s^*=z+\eta_d^*=1.34$, and $\omega=1.5$. The prefactors $a$ are 2, 2, and 4 for the quantity $\xi_m$, $\langle M^2_0\rangle$, and $\langle D^2_x\rangle$, respectively. Black dashed curves are  second-order curves  with corresponding critical exponents. }
 \label{fig:DataCollapse_J1y_02_055}
 \end{figure}

We now consider smaller $J_{1y}=0.6$, 0.55, and 0.2. At  $J_{1y}=0.6$, the QSL region shrinks to a narrow region $0.273\lesssim J_2 \lesssim 0.283$. At  $J_{1y}=0.55$ and 0.2, the QSL disappears and instead a direct AFM--VBS transition is suggested by the analysis of $\xi_m/L$ and $\langle D^2_x\rangle/ \langle M^2_0\rangle$. The results indicate a tricritical point between $J_{1y}=0.55$ and $J_{1y}=0.6$, which is roughly located at $(J_{1y},J_2)\simeq (0.58,0.27)$. These important results explicitly show how a gapless QSL emerges with $J_{1y}$ increasing. The data on collapse in Fig.\ref{fig:DataCollapse_J1y_065}--\ref{fig:DataCollapse_J1y_02_055} and critical exponents therein indeed support universality classes different from the class of the DQCP. Note that the spin and dimer correlation exponents $z+\eta_{s,d}$ at the QSL--VBS transition point have the same values, consistent with the emergent symmetry. 
 
 \subsection{C. Case that $J_{1y}=0.98$}
\begin{figure}[htbp]
 \centering
 \includegraphics[width=3.4in]{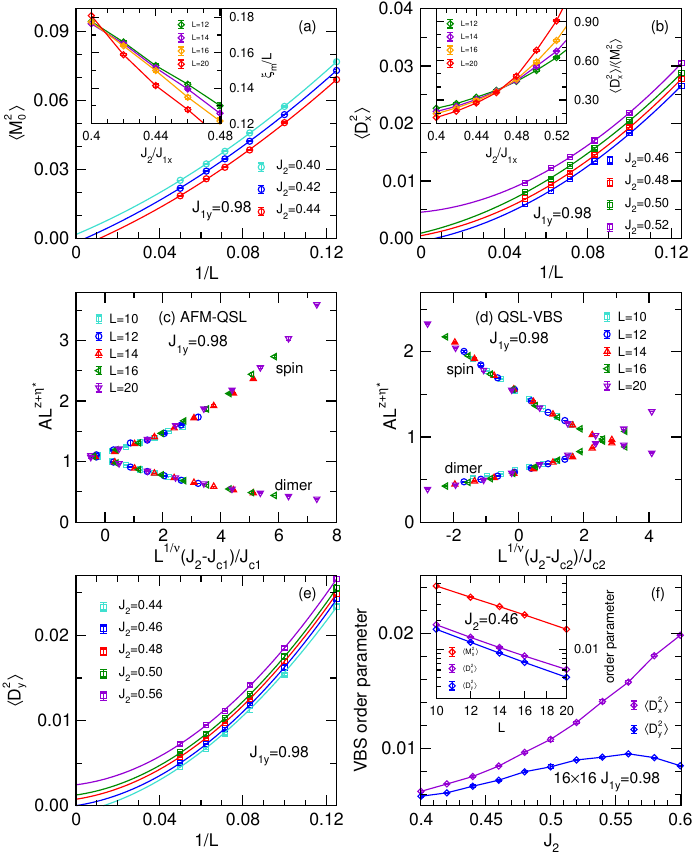}
 \caption{Physical behaviors at fixed $J_{1y}=0.98$. (a) Finite scaling of the AFM order parameter $\langle M^2_0\rangle $ (main panel) and spin correlation length $\xi_m$ (inset). (b) Finite size scaling of VBS order parameter $\langle D^2_x\rangle $  (main panel) and the order parameter ratio $\langle D^2_x\rangle/\langle M^2_0\rangle $  (inset). Second-order polynomial fits are used. (c) Data collapse at the AFM--QSL transition point $J_{c1}=0.41$, with critical exponents $z+\eta_{s1}=1.27$, $z+\eta_{d1}=1.86$, and $\nu=1.0$. (d) data collapse at the QSL--VBS transition point $J_{c2}=0.465$, with critical exponents $z+\eta_{s2}=z+\eta_{d2}=1.52$ and $\nu=1.0$. (e) Second-order fit of the VBS order parameter $\langle D^2_y\rangle $. (f) $J_2$-dependence of $\langle D^2_y\rangle $ on a $16\times 16$ lattice. The inset of (f) shows the order parameter scaling $A\propto L^{-\alpha}$ for $\langle M^2_0\rangle $, $\langle D^2_x\rangle $  and $\langle D^2_y\rangle $ at $J_2=0.46$. The exponents are $\alpha=$ 1.54(2), 1.57(3), and 1.70(2), respectively.}
 \label{fig:DataCollapse_J1y_098}
 \end{figure}

 The $J_{1x}$-$J_{1y}$-$J_2$ model is reduced to the $J_1$-$J_2$ model by setting $J_{1y}=1$, which was well studied in our previous work using the same method~\cite{liuQSL}. Therefore, we finally compute the case for fixed $J_{1y}=0.98$, as shown in Fig.\ref{fig:DataCollapse_J1y_098}. The finite size scaling of the AFM order parameter and correlation length quantities $\xi_m/L$ suggests that the AFM order vanishes at $J_{2}\simeq 0.41$. The QSL--VBS transition point estimated by the finite-size scaling of  $\langle D^2_x\rangle$  is  $J_2\simeq 0.47$, which is close to the crossing  value of  $\langle D^2_x\rangle/ \langle M^2_0\rangle$ (i.e., $J_2\simeq 0.465$). The critical exponents for AFM--QSL and QSL--VBS transitions obtained from the data collapse are well consistent with those obtained for other $J_{1y}$.  However, we find that $\langle D^2_y\rangle$ in the 2D limit is potentially nonzero for $J_2 > 0.46$. Note that $\langle D^2_y\rangle$ on a $16\times 16$ lattice at $J_{1y}=0.98$ has a behavior similar to that at $J_{1y}=0.95$, both having a peak in the VBS phase as seen in Fig.\ref{fig:DataCollapse_J1y_098}(f) and Fig.\ref{fig:VBS_stripe_J1y_095}. We plot the scaling behavior  of $\langle D^2_x\rangle$ and $\langle D^2_y \rangle$ at $J_2=0.46$, as shown in the inset of Fig.\ref{fig:DataCollapse_J1y_098}(f), and find that $\langle D^2_y \rangle$ decays more rapidly than $\langle D^2_x \rangle$. Therefore, we cannot exclude the possibility that nonzero values of $\langle D^2_y \rangle$ in 2D space are a finite-size effect.  In fact, when close to $J_{1y}=1.0$, there is an intermediate range of scale with approximate $C_4$ symmetry, which makes it challenging to get conclusive results in this situation. However, if it is not a finite-size effect, the nature of the VBS at $J_{1y}=0.98$ will be a mixed columnar-plaquette VBS phase where $\langle D^2_x \rangle$ and $\langle D^2_y \rangle$ have unequal nonzero values in the 2D limit.
 
Note that the lattice symmetry is $C_2$ for $J_{1y}\neq 1.0$ but $C_4$ for $J_{1y}=1.0$ (i.e., we have the $J_1$-$J_2$ model). It would be a little subtle to directly extend the anisotropic results to the $J_1$-$J_2$ model, especially for the VBS phase. Note that whether the nature of the VBS in the $J_1$-$J_2$ model is a columnar VBS (cVBS) phase or plaquette VBS (pVBS) phase in the thermodynamic limit is not clear. There are two possibilities assuming that the extension is continuous. (I) The VBS is in a cVBS phase in the $J_1$-$J_2$ model; this indicates that the VBS is also in the cVBS phase in the anisotropic case. The nonzero extrapolated $\langle D^2_y \rangle$ values  at $J_{1y}=0.98$ should then be finite-size effects. (II) The VBS is in the pVBS phase in the $J_1$-$J_2$ model, and to realize a continuous extension, there should be a mixed columnar--plaquette VBS phase that intervenes the cVBS and pVBS phases~\cite{didier2008}. In this situation, the $\langle D^2_y \rangle$ values in the VBS region  at $J_{1y}=0.98$ cannot be finite-size effects for a mixed columnar--plaquette VBS phase to be obtained. The two scenarios cannot be distinguished based on current capability.
 
 \subsection{D. Correlation functions} 
\begin{figure}[htbp]
 \centering
 \includegraphics[width=3.4in]{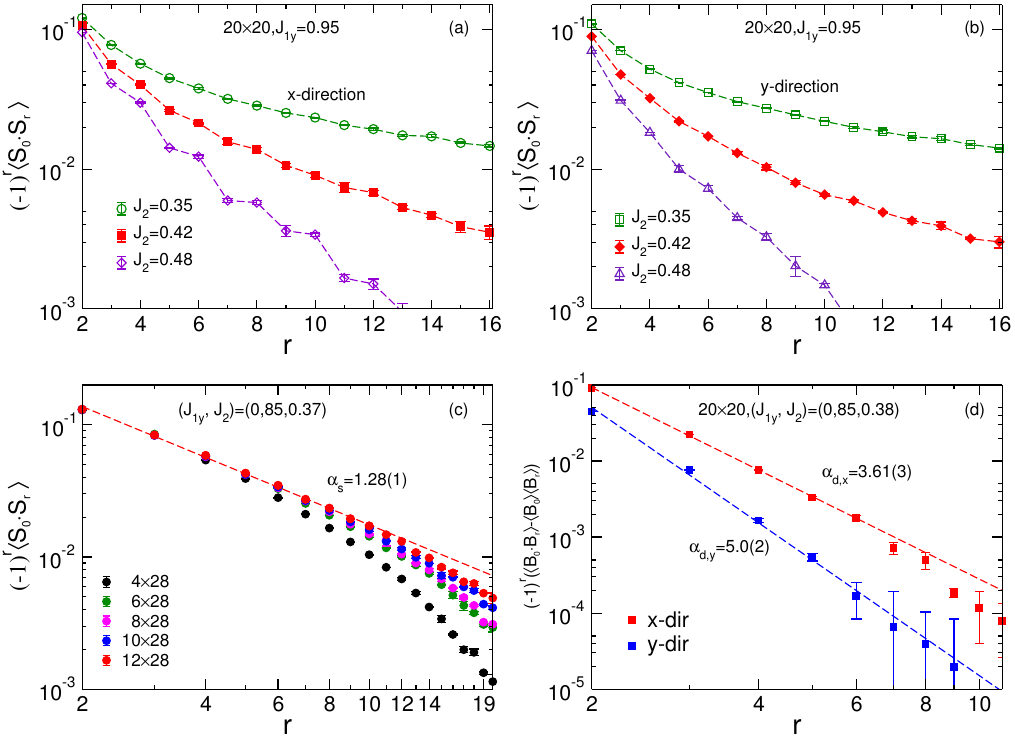}
 \caption{Spin and dimer correlation functions. Upper panels: (a) and (b) present the spin correlation functions along the $x$ direction on the central row and along the $y$ direction on the central column on the $20\times20$ lattice at different $J_2$ with fixed $J_{1y}=0.95$. Lower panels: (c) presents spin correlation functions along $x$ on the central row on the long strip $L_y\times L_x$ at $(J_{1y},J_2)=(0.85,0.37)$, with $L_y=4-12$ and $L_x=28$;(d) shows the dimer correlations along $x$ and $y$ directions on the $20\times 20$ lattice at $(J_{1y},J_2)=(0.85,0.38)$. Dashed lines show the power law fit $y=cr^{-\alpha}$ with $r\leq 10$ (6) for spin (dimer) power $\alpha_{s}$ ($\alpha_{d}$) as presented.}
 \label{fig:longstripJ2}
 \end{figure}

 In Fig.~\ref{fig:longstripJ2} (a) and (b), we show the spin correlation functions along $x$ and $y$ directions at $J_2=0.35$, 0.42, and 0.48 on a $20\times 20$ matrix with $J_{1y}=0.95$; these $J_2$ are within the regions of AFM, QSL, and VBS phases, respectively. We also show the spin correlation functions on long strips $L_y \times L_x$ with $L_y=4-12$ and $L_x=28$ at $(J_{1y},J_2)=(0.85,0.37)$, which is located within the region of the QSL phase.  The dimer correlation functions on a $20\times 20$ matrix for $(J_{1y},J_2)=(0.85,0.38)$ (in the region of the QSL phase) are shown in Fig.~\ref{fig:longstripJ2} (d).  These results  suggest that the QSL phase is gapless with a power law behavior for both spin and dimer correlation functions. Note that on the $20\times 20$ lattice, the correlations along $x$ and $y$ directions are different,  reflecting the anisotropy of the $x$ and $y$ directions in the finite size calculations.  Nevertheless, it is expected that the anisotropy in the  QSL phase will recover in the infrared limit. Note that without loss of generality, the distance between the reference site and the left edge used here is one lattice spacing for the $20\times 20$ lattice and three lattice spacings for the $12\times 28$ lattice. One can also choose other reference sites, such as those having two or four lattice spacings between the reference site and the left edge, and they also show a power law decay behavior of correlation functions.

 \section{3. Extracting critical exponents}

Generally speaking, the accurate computing of critical exponents is a challenging task in numerical simulations. Here, we have huge volumes of data with different $J_{1y}$, $J_2$, and size $L$, which makes it possible to extract meaningful critical exponents.
Following Ref.~\cite{liuj1j2j3}, we use the standard formula to collectively fit the physical quantities from different lattice sizes and different couplings for data collapse:
 \begin{equation}
A(J_2, L)=L^{\kappa}(1+aL^{-\omega} )F[L^{1/\nu}(J_2-J_c)/J_c],
\end{equation}
 where $A= \xi_m$, $\ave{M^2_0}$, or  $\ave{D^2_x}$, and $\kappa=1$ for $ A= \xi_m$, $-(z+\eta_s^*)$ for $ A= \ave{M^2_0}$, and  $-(z+\eta_d^*)$ for $ A= \ave{D^2_x}$. Factors $a$ and $\omega$ are tuning parameters of the subleading term.  $F[]$ is a polynomial function, and we here use a third-order expansion and find that the second-order fit already works well because the third-order coefficient is small. Generally, for the AFM--VBS transition, the subleading term is necessary for good data collapse, although this is not always the case, such as when $J_{1y}$=0.4.  For AFM--QSL and QSL--VBS transitions, subleading terms seem unnecessary, and we set $a=0$.  The transition point $J_c$ is always fixed unless otherwise specified. In the following, we focus on how to evaluate the critical exponents for AFM--QSL and QSL--VBS transitions, and similar analyses can be conducted for AFM--VBS transitions.
 
 For a given $J_{1y}$, we can estimate the AFM--QSL transition point $J_{c1}$ using the crossing of $\xi_m/L$. Taking account of possible finite-size effects, we can alternatively collectively fit  $J_{c1}$ and $\nu_1$ simultaneously using the values of $\xi_m/L$ for different $J_2$ and $L$ according to the above formula. Suppose that at $J_{1y}=0.85$ we have $J_{c1}=0.352(5)$ and $\nu_{1m}=1.00(4)$. With this fixed $J_{c1}=0.352$, we can then use a collective fit of  $\nu$ and $z+\eta^*$ for the AFM and VBS order parameters, respectively. We thus obtain $\nu_{1,s}=1.00(3)$ and spin correlation exponent $z+\eta_{s1}^*=1.18(1)$ by fitting the AFM order parameters, and $\nu_{1,d}=0.98(2)$ and dimer correlation exponent $z+\eta_{1,d}^*=1.86(2)$ by fitting the VBS order parameters. At the QSL--VBS transition, we mention again that the dimer correlation length cannot be obtained to locate the VBS phase boundary for open-boundary systems because the dimer structure factor in this case is not well defined~\cite{zhao2020,liuQSL}. We here  use the critical point $J_{c2}$ obtained using the order parameter ratio $\ave{D^2_x}/\ave{M^2_0}$, which has very small finite-size effects. Through fixing $J_{c2}$, we similarly get $\nu_{2,s}=0.97(5)$ and spin correlation exponent $z+\eta_{s2}^*=1.50(1)$ by fitting AFM order parameters, and $\nu_{1,d}=1.04(3)$ and dimer correlation exponent $z+\eta_{2,d}^*=1.50(1)$ by fitting VBS order parameters.  
 
We apply the same analyses to other cases of a fixed $J_{1y}$. With given $J_{c1}$ obtained from $\xi_m$ and $J_{c2}$ obtained from $\ave{D^2_x}/\ave{M^2_0}$, by scaling order parameters, we get $\nu_{1,s}$, $\nu_{1,d}$,  $\nu_{2,s}$, and $\nu_{2,d}$ and their corresponding $z+\eta_{s1}^*$, $z+\eta_{d1}^*$, $z+\eta_{s2}^*$, and $z+\eta_{d2}^*$.
Table~\ref{tab:freenu} lists the fitted exponents $\nu_{1,2}$ and $\nu_{1m}$ obtained by fitting $\xi_m$ at the AFM--QSL transition point,  as well as the spin and dimer correlation exponents $z+\eta_{s,d}^*$. Note that the obtained $\nu_1$ and $\nu_2$ are close, indicating $\nu_1 \approx \nu_2\approx 1.0$. Additionally, the spin and dimer correlation exponents are well consistent, $z+\eta_{s1}^*\sim 1.2$ and $z+\eta_{d1}^*\sim 1.8$ at the AFM--QSL transition, and $z+\eta_{s2}^*=z+\eta_{d2}^*\sim 1.5$ at the QSL--VBS transitions.

 \begin{table}
    \centering
 \caption { Fitted critical exponents and errors from fittings.  Fitting AFM order parameters gives $\nu_{1,s}$ and  $z+\eta_{s1}^*$ at the AFM--QSL critical point, and gives $\nu_{2,s}$ and  $z+\eta_{s2}^*$ at the QSL--VBS critical point. Fitting VBS order parameters gives $\nu_{1,d}$ and  $z+\eta_{d1}^*$ at the AFM--QSL critical point, and gives $\nu_{2,s}$ and  $z+\eta_{d2}^*$ at the QSL--VBS critical point.  The last column $\bar{\nu}$ is an estimation made by directly averaging over $\nu_{1,s}$, $\nu_{1,d}$,  $\nu_{2,s}$, $\nu_{2,d}$, and $\nu_{1m}$.}
  	\begin{tabular*}{\hsize}{@{}@{\extracolsep{\fill}}lcccccc@{}}
		\hline\hline
		        &  $\nu_{1,s}$ & $\nu_{1,d}$   & $\nu_{2,s}$& $\nu_{2,d}$& $\nu_{1m}$   & $\bar{\nu}$ \\ \hline
		      $J_{1y}=0.60$& 0.94(2)& 0.91(4)  & 0.95(2) & 0.93(3) &1.11(5)& 0.97(4)   \\ \hline
		      $J_{1y}=0.65$& 1.06(3) & 0.95(5)   &  1.02(5) & 0.93(3)&1.05(7)& 1.00(5)   \\ \hline
	          $J_{1y}=0.75$& 1.02(5) & 0.96(3)   &  1.01(4) & 0.97(5)& 1.07(4)& 1.01(4)  \\ \hline 
	     $J_{1y}=0.85$& 1.00(3) & 0.98(2)   &  0.97(5) & 1.04(3)& 1.00(5) & 1.00(4)\\ \hline
  	     $J_{1y}=0.95$& 1.03(5) & 1.07(8)   &  1.05(3) & 1.10(6)& 1.01(4) & 1.05(5)\\ \hline
	        &  $z+\eta_{s1}^*$ & $z+\eta_{d1}^*$   & $z+\eta_{s2}^*$& $z+\eta_{d2}^*$   \\ \hline
	             $J_{1y}=0.60$& 1.35(4) & 1.55(2)   & 1.45(2)  & 1.45(1)  \\ \hline
	        	     $J_{1y}=0.65$& 1.20(1) & 1.70(1)   &  1.49(1) & 1.48(1)   \\ \hline
	          $J_{1y}=0.75$& 1.23(1) & 1.76(1)   &  1.47(1) & 1.47(1) \\ \hline 
	     $J_{1y}=0.85$& 1.18(1) & 1.86(2)   &  1.50(1) & 1.50(1)  \\ \hline
  	      $J_{1y}=0.95$& 1.21(2) & 1.87(2)   &  1.52(1) & 1.53(2) \\ \hline
	   		\hline\hline
	\end{tabular*}
\label{tab:freenu}	
\end{table}

 The closeness of $\nu_1$ and $\nu_2$ indicates that the AFM--QSL and QSL--VBS transitions could have the same correlation length exponent $\nu$ and a single $\nu$ could scale all the quantities.  To demonstrate this point, we use an value $\bar{\nu}$ averaged over  $\nu_{1,s}$, $\nu_{1,d}$,  $\nu_{1m}$, $\nu_{2,s}$, and $\nu_{2,d}$ as a fixed parameter to fit $z+\eta^*$.  The data collapse with the single $\bar{\nu}$ for all the cases was shown in previous sections. Corresponding exponents are listed in Table~\ref{tab:fixnu} and only slightly differ from those in Table~\ref{tab:freenu}, which means that a single $\nu\sim 1.0$  indeed works well at the AFM--QSL and QSL--VBS transitions. We note that due to the imperfect optimization of wave functions, some obtained physical quantities may have slight unavoidable deviations from their exact values.  However, this would not diminish the reasonability and correctness of the extracted critical exponents, because the large volumes of data from different $(J_{1y},J_2)$ reduce the uncertainty. Note that the physical quantities at different $J_{1y}$ can be  scaled well using smooth curves with similar critical exponents, which is an excellent characterization of  the universal scaling functions.
 \begin{table}
    \centering
 \caption { Critical exponents obtained using a single correlation length exponent $\bar{\nu}$ at the AFM--QSL and QSL--VBS transition points.  In each fit, $z+\eta^*$ is a free parameter, and  $\bar{\nu}$ is fixed. Errors are from fittings. }
  	\begin{tabular*}{\hsize}{@{}@{\extracolsep{\fill}}lccccc@{}}
		\hline\hline
	        &  $z+\eta_{s1}^*$ & $z+\eta_{d1}^*$   & $z+\eta_{s2}^*$& $z+\eta_{d2}^*$ & $\bar{\nu}$  \\ \hline
       	          $J_{1y}=0.60$& 1.36(1) & 1.55(1)   & 1.44(2)   & 1.45(1)  & 0.97(4) \\ \hline 
	          $J_{1y}=0.65$& 1.23(2) & 1.70(1)   &  1.49(1) & 1.49(1) & 1.00(5)\\ \hline 
	          $J_{1y}=0.75$& 1.23(1) & 1.74(1)   &  1.47(1) & 1.47(1) & 1.01(4)\\ \hline 
	          $J_{1y}=0.85$& 1.18(1) & 1.85(2)   &  1.50(1) & 1.50(2) & 1.00(4)\\ \hline 
	          $J_{1y}=0.95$& 1.21(2) & 1.88(2)   &  1.52(1)& 1.52(2) & 1.05(5)\\ \hline 
	\end{tabular*}
\label{tab:fixnu}	
\end{table}

\bibliography{J1xJ1yJ2}
\end{document}